\documentclass [12pt]{article}
\usepackage{amssymb}
\usepackage{amsmath}
\usepackage{graphicx}
\usepackage{amscd}
\usepackage {longtable}
\usepackage{epsfig}
\date{}

\title{QUADRUPOLE OSCILLATIONS AS PARADIGM OF THE CHAOTIC MOTION IN
NUCLEI. Part 2}

\author{V.P.Berezovoj\footnote{e-mail: berezovoj@kipt.kharkov.ua},
 Yu.L. Bolotin, V. Yu. Gonchar,
${}^{\star}$M.Ya.Granovsky,\\ V.N. Tarasov}

\begin{document}

\maketitle

\begin{center}{\it
National Science Center "Kharkov Institute of Physics and Technology", Kharkov, 310108, Ukraine\\

${}^\star$Pro Training Tutorial Institute, 18 Stoke Av, Kew 3101,
Melbourne, Vic., Australia }

\vspace{1cm}
\end{center}
\setcounter{figure}{17}

The present communication is a continuation of the review, the
first part of which was presented in nucl-th\ 0109033. The second
part dealswith the manifistation of Chaotic dynamics at quantum
level. The variations of statistical properties of energy spectrum
in the process of $R - C - R$ transition have been studied in
detail. We proved that the type of the classical motion is
correlated with the structure of the eigenfunctions of highly
excited states in the $R - C - R$ transition. Shell structure
destruction induced by the increase of nonintegrable perturbation
was analyzed.

\vspace{2cm}


\newpage

\tableofcontents

\section {Quantum manifestations of the classical \\stochasticity}

\subsection{ The quantum chaos problem}

Essential progress in the understanding of the nonlinear dynamics
of classical systems stimulated numerous attempts to include the
conception of the stochasticity in quantum mechanics
\cite{4,6,53,54,55,56,57}. The essence of the problem consists in
the fact that the energy spectrum of any quantum system,
performing the finite motion, is discrete and therefore its
evolution is quasiperiodic. At the same time, the correspondence
principle requires the possibility of limit transition to
classical mechanics, containing not only regular solutions but
also chaotic ones. The total solving of this problem is too
complicated one, therefore its restricted variant is of
considerable interest. We are interested in the search of
peculiarities of behaviour of quantum systems the classical
analogs of which reveal chaotic behaviour. Such peculiarities are
usually called the quantum manifestations of the classical
stochasticity (QMCS).

The following objects can be used as the objects of search of the
QMCS for an autonomous systems

1) the energy spectra

2) the stationary wave functions

3) the wave packets.

A priori the manifestations of the QMCS can be expected both in
the form of some peculiarities of concrete stationary state and in
the whole group of close states in energy. Of course, it is not
excepted the possibility that such alternative does not exist at
all, i.e. the traces of the classical chaos can be detected both
in the properties of separate states and so in their sets.

In this part of the review we will represent the results relating
to the QMCS in the dynamics of quadrupole oscillations.

\subsection{ Numerical procedure}

The quantum scaled Hamiltonian describing the quadrupole surface
oscillations has the following form
\begin{equation}\label{3.2.1}
H = \frac{{\bar \hbar _i^2 }} {2}\left( {\Delta _x  + \Delta _y }
\right) + U_i \left( {x,y,W} \right)
\end{equation}
where the scaled Planck's constant is $ \bar \hbar  =
{\raise0.7ex\hbox{$\hbar $} \!\mathord{\left/
 {\vphantom {\hbar  {\hbar _{0i} }}}\right.\kern-\nulldelimiterspace}
\!\lower0.7ex\hbox{${\hbar _{0i} }$}} $ , $ \hbar _{0i}^2  =
m\varepsilon _{0i} l_{0i}^2$ , and scaling parameters $\hbar _{0i}
,l_{0i} $ and deformation potential $U_i (x,y,W)$ are determined
by the expressions (21.1) and (23.1). We are going to study the
peculiarities of structure of energy spectra and wave functions in
each of the following intervals

the first regular region $R_1 :\quad 0 < E < E_{cr1} $

the chaotic region          $C:\quad E_{cr1}  < E < E_{cr2} $

the second regular region        $R_2 :\quad E > E_{cr2} $.

The critical energies $E_{cr1} $ and $E_{cr2} $ were determined
above. The main numerical calculations were performed for one-well
potentials ($0 < W < 16$ ). In this case the critical energies are
determined by the relations (60.1), (61.1).

At fixed topology of potential surface $\left( {W = const}
\right)$ , the scaled Planck's constant $ \bar \hbar _i$ is the
unique free parameter in the quantum Hamiltonian (\ref{3.2.1}). In
the study of the concrete region of energies, corresponding to the
certain type of classical motion $\left( {R_1 ,C,R_2 } \right)$ ,
the choice of $ \bar \hbar _i $ is dictated by the possibility of
attainment the required degree of validity of semiclassical
approximation (large quantum numbers) conserving the precision of
calculations of spectrum and wave functions (the limitation of the
possibility of diagonalization of large dimension matrices). For
the original nonreduced Hamiltonian (16.1) the variation of $\bar
\hbar _i$ is equivalent to the variation of the critical energies
$E_{cr1} $ and $E_{cr2} $ (i.e. parameters $a,b,c$ ) or to the
adequate choice of Planck's constant $ \hbar$ .

The procedure of diagonalization was used for the determination of
energy spectrum and eigenfunctions of the Hamiltonian
(\ref{3.2.1}). As the basis we chose the simple combinations of
the eigenfunctions of the two-dimensional harmonic oscillator with
equal frequencies \cite{58}
\begin{equation}\label{3.2.2}
\begin{gathered}
  \left| {NLj} \right\rangle  = \frac{{P_{L,j} }}
{{\sqrt 2 }}\left( {\left| {NL} \right\rangle  + j\left| {N, - L} \right\rangle } \right),j =  \pm 1\; \hfill \\
  N = 0,1,2...;\quad L = N,N - 2...1\;\;or\;\;0 \hfill \\
\end{gathered}
\end{equation}
normalized according to the condition
\begin{equation}\label{3.2.3}
  \left\langle {{NLj}}
 \mathrel{\left | {\vphantom {{NLj} {N'L'j'}}}
 \right. \kern-\nulldelimiterspace}
 {{N'L'j'}} \right\rangle  = \frac{1}
{{\sqrt 2 }}2^{\delta _{L,0} } \delta _{jj'} \delta _{NN'} \delta
_{LL'}
\end{equation}
and $ P_{L,j}  = j^{Mod\left( {L,3} \right)}$ .

The symmetry of the considered Hamiltonian $(C_{3v} )$ leads to
the block structure of matrix $ \left\langle {N'L'j'}
\right|H\left| {NLj} \right\rangle$ , which consists of four
independent submatrices. It allows to carry out diagonalization of
each submatrix separately, obtaining in this case four independent
sets of states, which according to the classification used in the
theory of groups are the following states

$A_1 $ - type [mode$(L,3) = 0,j = 1$ , including $L = 0$ ]

$A_2 $ - type [mode$(L,3) = 0,j =  - 1,L \ne 0$ ]

and twice degenerated states

$E$ - type [ mode $(L,3) \ne 0,j =  \pm 1$ ]

The possibility of separate diagonalization of submatrices of
definite type allows essentially to increase available dimension
of basis in numerical calculations (later for brief, we shall call
the dimension of basis as the dimension n of separate submatrix).
Table 3.2.1 shows the maximum oscillator numbers$N_0 $ of the
basis and the dimension $m$ of total matrix $ \left\langle
{N'L'j'} \right|H\left| {NLj} \right\rangle$ for the above
enumerated types of submatrices with the dimension $n = 408$
\begin{table}
\begin{center}
Table 3.2.1\\
\medskip
\begin{tabular}{|c|c|c|c|} \hline
   & $A_1 $ & $A_2 $ & $E_{1,2} $ \\ \hline
  $N_0 $ & $67$ & $70$  & $48$ \\ \hline
  $m$ & $2346$ & $2556$ & $1225$ \\ \hline
\end{tabular}
\end{center}
\end{table}
\begin{center}
\end{center}

Expansion of wave eigenfunctions in basis (\ref{3.2.2}) in polar
coordinates $r,\varphi $
\begin{equation}\label{3.2.4}
\left\langle {{r,\varphi }}
 \mathrel{\left | {\vphantom {{r,\varphi } {E_k }}}
 \right. \kern-\nulldelimiterspace}
 {{E_k }} \right\rangle  = \sum\limits_{NL} {C_{NLj}^{\left( k \right)} \left\langle {{r,\varphi }}
 \mathrel{\left | {\vphantom {{r,\varphi } {NLj}}}
 \right. \kern-\nulldelimiterspace}
 {{NLj}} \right\rangle }
\end{equation}
where
\begin{eqnarray}\label{3.2.5}
&&\left\langle {{r,\varphi }}
 \mathrel{\left | {\vphantom {{r,\varphi } {NL}}}
 \right. \kern-\nulldelimiterspace}
 {{NL}} \right\rangle  = \\&&i^N \frac{{e^{ - iL\varphi } }}
{{\sqrt {2\pi } }}\frac{1} {{L!}}\left[ {\frac{{2\left( {\frac{{N
+ L}} {2}} \right)!}} {{\left( {\frac{{N - L}} {2}} \right)!}}}
\right]^{{\raise0.7ex\hbox{$1$} \!\mathord{\left/
 {\vphantom {1 2}}\right.\kern-\nulldelimiterspace}
\!\lower0.7ex\hbox{$2$}}} \left( {\sqrt {\omega _0 } r} \right)^L
e^{ - \omega _0 {\hbox{${r^2 }$} \!\mathord{\left/
 {\vphantom {{r^2 } 2}}\right.\kern-\nulldelimiterspace}
\!\lower0.7ex\hbox{$2$}}} M\left\{ { - \frac{{N - L}} {2},L +
1,\omega _0 r^2 } \right\}\nonumber
\end{eqnarray}
Here $M\{ \quad \} $ is degenerated hypergeometric function and
$\omega _0 $ - frequency of the oscillator basis. The finite
dimension of the basis used in calculations leads to the
dependence of the position of level from the oscillator frequency
$\omega _0 $ . Such dependence for the dimensions of basis $n =
198$ and $n = 408$ ($W = 13$ ) is represented in Fig.18.
\begin{figure}
\centering
\includegraphics[height=8cm]{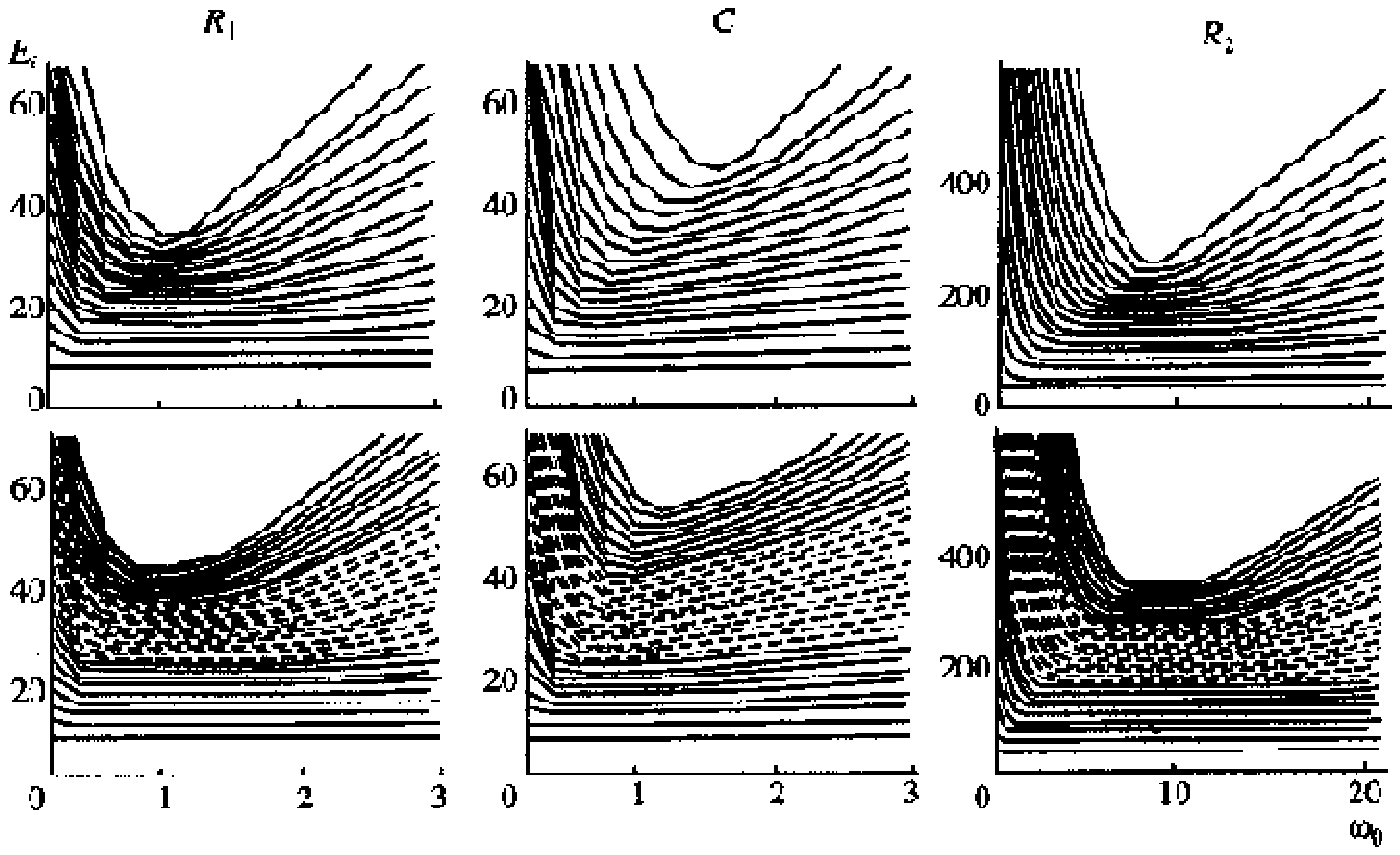}
\caption{Energy spectra of Hamiltonian (\ref{3.2.1}) as function
of the frequency $\omega _0 $ of the oscillator basis $\left( {W =
13} \right)$ . The upper and lower parts of this figure show this
dependence for the basis dimensions $n = 198$ and $n = 408$ ,
respectively.} \label{Fig.18}
\end{figure}

The important moment of numerical calculations is the choice of
the optimal frequency $\omega _0n ^{opt} $, determined from the
condition that the energies of levels, situated in the interval of
our interest, have minimum value. Such optimization is essentially
important for the regions $C$ and $R_2 $ , where the position of
the levels more depends on the frequency of basis. The optimal
frequency depends both on dimension of the basis and on the length
of investigated energy interval. As is shown in Fig.18 it is
possible to choose the unique frequency $\omega _0 ^{opt} $ for
intervals including decades or more levels just at the dimension
of the basis $n = 408$ . These levels are selected in Fig.18 by
dotted lines. The large  value of $\omega _0 ^{opt} $ for the
region $R_2 (\omega _0 ^{opt} \sim 8)$ is caused by the fact that
at energies $E > E_{cr2} $ the deformation potential essentially
differs from harmonic oscillator potential.

The dimension of the basis used in our calculations was determined
for the reason of acceptability of the calculation time and the
calculation precision of the position of levels in considered
region. The results of investigation of saturation in basis are
represented in Fig.19, i.e. dependence of the results of spectrum
calculation from the dimension of basis. As it is shown in Fig.19
for the regions $R_1 $ and $R_2 $ acceptable precision of
calculations of levels with ordinal numbers $100 - 200$ is reached
at the dimension of the basis $n\sim 400$ , while for the chaotic
region $C$ dimension of the basis required for the attainment of
the same precision considerably larger $(n\sim 700)$ . It is a
well known result \cite{4,12}: the value of saturation in basis is
sensitive to the type of classical motion.
\begin{figure}
\centering
\includegraphics[height=14cm]{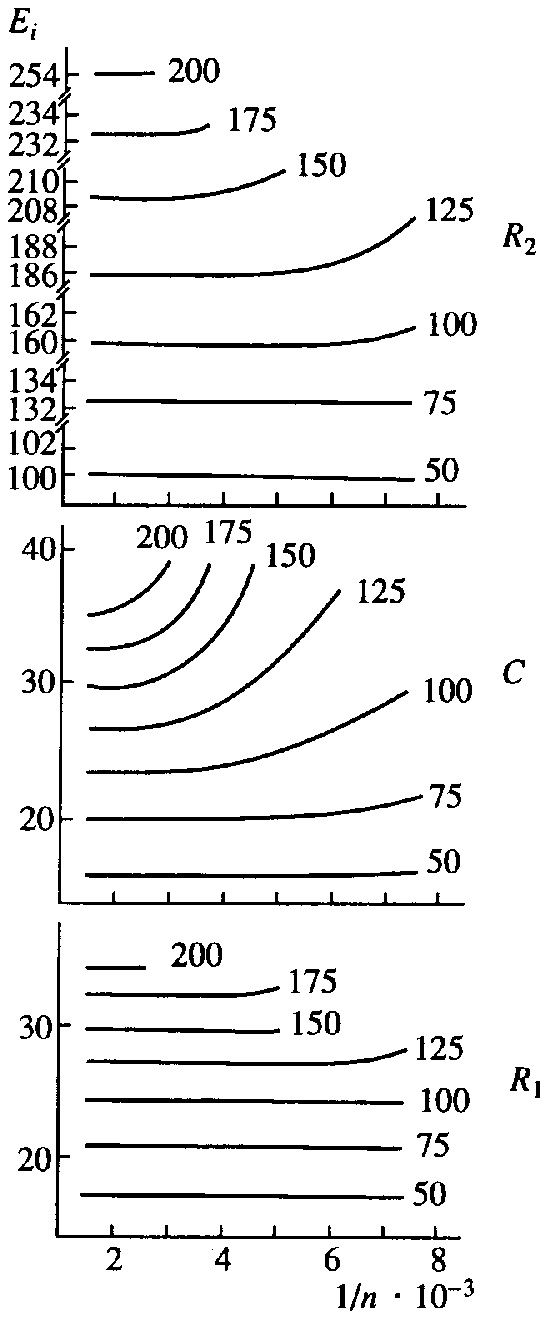}
\caption{Energy levels as function of the basis dimension $n$ .
The numbers on the curves are the level numbers $k$ . }
\label{Fig.19}
\end{figure}

Matrix diagonalization is attractive for treating Hamiltonians
that do not differ greatly from Hamiltonians with known
eigenfunctions. This numerical procedure becomes less attractive
(or even not effective at all) at the transition to the PES of
complicated topology, when required Hamiltonian matrix is of high
order. In this case the alternative to the diagonalization may
become the so-called spectral method \cite{59}, which utilizes
numerical solutions to the time-dependent Schrodinger equation.
The spectral method was developed earlier for determining the
eigenvalues and eigenfunctions for the modes of optical waveguides
from numerical solutions of the paraxial wave equation \cite{60}.
Feit, Fleck and Steiger \cite{59}  were able to apply the
previously developed methodology to quantum mechanical problems
with little change, as the latter equation is identical to the
Schrodinger equation.

The spectral method requires computation of the correlation
function
\begin{equation}\label{3.2.6}
 P_1 \left( t \right) = \left\langle {{\psi \left( {\vec r,0} \right)}}
 \mathrel{\left | {\vphantom {{\psi \left( {\vec r,0} \right)}
 {\psi \left( {\vec r,t} \right)}}}
 \right. \kern-\nulldelimiterspace}
 {{\psi \left( {\vec r,t} \right)}} \right\rangle
\end{equation}
where $ \psi \left( {\vec r,t} \right) $ represents a numerical
solution of the time-dependent Schrodinger equation, and $ \psi
\left( {\vec r,0} \right) $ is the wave function at $t = 0$ . The
solution $ \psi \left( {\vec r,t} \right) $ can be accurately
generated with the help of the split operator FFT (Fast Fourier
Transformation) method \cite{61}. The numerical FFT of $P_1 (t)$ ,
or $P_1 (E)$ , displays a set of sharp local maxima for $E = E_n $
, where $E_n $ are the desired energy eigenvalues. With the aid of
lineshape fitting techniques, both the positions and heights of
these resonances can be determined with high accuracy. The former
yield the eigenvalues and the latter the weights of the stationary
states that compose the wave packet once the eigenvalues are
known, the corresponding eigenfunctions can be computed by
numerically evaluating the integrals
\begin{equation}\label{3.2.7}
 \psi \left( {\vec r,E_n } \right) = \int\limits_0^T
 {\psi \left( {\vec r,t} \right)W\left( t \right)\exp \left( {iE_n t} \right)dt}
\end{equation}
where $T$ is the time encompassed by the calculation, and $W(t)$
is a window function,
\begin{equation}\label{3.2.8}
\begin{gathered}
  W\left( t \right) = 1 - \cos {\raise0.7ex\hbox{${2\pi t}$} \!\mathord{\left/
 {\vphantom {{2\pi t} T}}\right.\kern-\nulldelimiterspace}
\!\lower0.7ex\hbox{$T$}}\;\;if\quad 0 < t < T \hfill \\
  \quad \quad  = 0\quad \quad \quad \quad \quad if\quad \quad \;\;t > T \hfill \\
\end{gathered}
\end{equation}
Since the spectral method is fundamentally based on numerical
solutions of a time-dependent differential equation, its
implementation is always straightforward. No special ad hoc
selection of basis function is required, nor is it necessary for
the potential to have a special analytic form. The spectral method
is in principle applicable to problems involving any number of
dimensions.

\subsection{ Quantization by the normal form}

In this section we calculate a semiclassical approximation to an
energy spectrum of the Hamiltonian of qudrupole oscillations
(\ref{3.2.1}) \cite{62} (here it will be more suitable for us to
use the nonscale version of the Hamiltonian (16.1) with $m = 1$ )
by quantization the normal form and compare results to exact
quantum mechanical calculations. By the exact spectrum we mean the
spectrum obtained by the direct numerical calculations, for
example, by the diagonalization of the Hamiltonian on the
reasonably chosen basis. Quantization the incommensurable case is
straightforward. Since the normal form can be expressed entirely
in terms of action variables, we merely have to replace those
action variables by an appropriate multiple of  $ \hbar ,I_\nu \to
\hbar \left( {n_\nu   + {\raise0.7ex\hbox{$1$} \!\mathord{\left/
 {\vphantom {1 2}}\right.\kern-\nulldelimiterspace}
\!\lower0.7ex\hbox{$2$}}} \right) $ and the result is a power
series in the quantum numbers,
\begin{equation}\label{3.3.1}
  E_{n_1 n_2 }  = \Gamma \left( {n_1  + {\raise0.7ex\hbox{$1$} \!\mathord{\left/
 {\vphantom {1 2}}\right.\kern-\nulldelimiterspace}
\!\lower0.7ex\hbox{$2$}},n_2  + {\raise0.7ex\hbox{$1$}
\!\mathord{\left/
 {\vphantom {1 2}}\right.\kern-\nulldelimiterspace}
\!\lower0.7ex\hbox{$2$}}} \right)
\end{equation}
The commensurable case is less straightforward. Commensurability
of frequencies leads to the appearance of angle variables in the
normal form
\begin{equation}\label{3.3.2}
\Gamma  = \Gamma \left( {I_1 ,I_2 ,\varphi _2 } \right)
\end{equation}
As before, the action $I_1 $ we can replace on $ \hbar \left( {n_1
+ {\raise0.7ex\hbox{$1$} \!\mathord{\left/
 {\vphantom {1 2}}\right.\kern-\nulldelimiterspace}
\!\lower0.7ex\hbox{$2$}}} \right) $ . In this case the initial
two-dimensional problem is reduced to one-dimensional problem, the
quantization of which can be performed with the help of WKB
method.

The procedure of quantization by normal form we shall begin with
canonical transformation
\begin{equation}\label{3.3.3}
  \begin{gathered}
  q_1  = \frac{i}
{2}\left( { - Q_1  + Q_2  + P_1  - P_2 } \right) \hfill \\
  q_2  = \frac{1}
{2}\left( {Q_1  + Q_2  + P_1  + P_2 } \right) \hfill \\
  p_1  = \frac{1}
{2}\left( {Q_1  - Q_2  + P_1  - P_2 } \right) \hfill \\
  p_2  = \frac{i}
{2}\left( {Q_1  - Q_2  - P_1  - P_2 } \right) \hfill \\
\end{gathered}
\end{equation}
where variables $(\vec p,\vec q)$ provide reduction of the
harmonic part of the initial Hamiltonian (16.1) to the form
(77.1). In the new variables $\vec Q(Q_1 ,Q_2 )$ and $\vec P(P_1
,P_2 )$ the Hamiltonian of quadrupole oscillations is the
following
\begin{equation}\label{3.3.4}
 K\left( {\vec Q,\vec P} \right) = K^{\left( 2 \right)} \left(
 {\vec Q,\vec P} \right) + \sum\limits_{j > 2} {K^{\left( j
  \right)} \left( {\vec Q,\vec P} \right)}
\end{equation}
where
\begin{equation}\label{3.3.5}
  K^{\left( 2 \right)} \left( {\vec Q,\vec P} \right) =
  i\left( {Q_1 P_1  + Q_2 P_2 } \right)
\end{equation}
and $K^{(j)} $ are homogeneous polynomials in variables $Q_i $ and
$P_i $ of degree $j$. Each member $K^{(j)} $ of the Hamiltonian
(\ref{3.3.4}) is reduced to the normal form according to the
procedure describing in the section 2.5(part 1). However this
procedure is sufficiently simplified due to the diagonal form of
the operator $D$ in the variables $(\vec P,\vec Q)$. The classical
normal form is the sum of the polynomials of special form in $Q_i
$ and $P_i $ . For obtaining its quantum analog we can use the
Weyl's heuristic rule of correspondence \cite{63,64}
\begin{equation}\label{3.3.6}
P^n Q^m  = Q^m P^n  \to \frac{1} {{2^n }}\sum\limits_{l = 0}^n
{\frac{{n!}} {{l!\left( {n - l} \right)!}}} \hat Q^m \hat P^{n -
l}
\end{equation}
The operators $\hat P_i $ and $\hat Q_i $ are determined by
formulae (\ref{3.3.3}), in which by $p_i $ and $q_i $ we should
mean operators of impulse and coordinate with usual rules of
commutation, from which it follows that
\begin{equation}\label{3.3.7}
\left[ {\hat P_k ,\hat Q_l } \right] = \delta _{kl} \left( {k,l =
1,2} \right)
\end{equation}
The operators $\hat P_i $ , $\hat Q_i $ allow to introduce the
full orthonormalized basis
\begin{equation}\label{3.3.8}
\left| {NL} \right\rangle  = \left[ {\left( {\frac{{N + L}} {2}}
\right)!\left( {\frac{{N - L}} {2}} \right)!} \right]^{ - \frac{1}
{2}} \hat Q_2^{\frac{{N - L}} {2}} \hat Q_1^{\frac{{N + L}} {2}}
\left| 0 \right\rangle
\end{equation}
where vacuum state $ \left| 0 \right\rangle $ is determined by
\begin{equation}\label{3.3.9}
  \hat P_1 \left| 0 \right\rangle  = \hat P_2 \left| 0 \right\rangle  = 0
\end{equation}
The principal quantum number $N = 0,1,...$ and angular momentum
number $L$ at given $N$ is equal $ \pm N, \pm (N - 2)...0$ or $1$.
The action of the operators $ \hat Q_i$ and $ \hat P_i $ on the
basis $ \left| {NL} \right\rangle$ is
\begin{eqnarray}\label{3.3.10}
\hat Q_1 \left| {NL} \right\rangle \!\! &=& \!\!\sqrt {\frac{{N +
L + 2}} {2}} \left| {N + 1,L + 1} \right\rangle , \hat P_1 \left|
{NL} \right\rangle  = \sqrt {\frac{{N + L}} {2}} \left| {N - 1,L -
1} \right\rangle\\
\hat Q_2 \left| {NL} \right\rangle  &=& \sqrt {\frac{{N - L + 2}}
{2}} \left| {N + 1,L - 1} \right\rangle ,\;\hat P_2 \left| {NL}
\right\rangle  = \sqrt {\frac{{N - L}} {2}} \left| {N - 1,L + 1}
\right\rangle\nonumber
\end{eqnarray}
Note that the constructed basis is oscillator one
\begin{equation}\label{3.3.11}
 \hat K^{\left( 2 \right)} \left| {NL} \right\rangle  =
  \left( {\hat Q_1 \hat P_1  + \hat Q_2 \hat P_2  + 1}
  \right)\left| {NL} \right\rangle  = \left( {N + 1}
  \right)\left| {NL} \right\rangle
\end{equation}
The classical normal form for considering Hamiltonian is (through
fourth degree)
\begin{multline}\label{3.3.12}
\Gamma  = \Gamma ^{\left( 2 \right)}  + \Gamma ^{\left( 4 \right)}
=\\
   i\left[ {\left( {Q_1 P_1  + Q_2 P_2 } \right) + {\raise0.7ex\hbox{${b^2 }$}
  \!\mathord{\left/
 {\vphantom {{b^2 } 6}}\right.\kern-\nulldelimiterspace}
\!\lower0.7ex\hbox{$6$}}\left( {P_1^2 Q_1^2  + P_2^2 Q_2^2
 - 12Q_1 Q_2 P_1 P_2 } \right) + } \right. \\
  \left. { + c\left( {P_1^2 Q_1^2  + P_2^2 Q_2^2  + 4Q_1 Q_2 P_1 P_2 }
  \right)} \right]
\end{multline}
The quantum normal form $ \hat \Gamma $ reconstructs from the
classical one with the help of Weyl's rule (\ref{3.3.6}) and
Dirac's rule of correspondence $ \hat \Gamma  \to \frac{1}
{i}\Gamma $
\[\hat \Gamma  = \left( {\hat Q_1 \hat P_1  + \hat Q_2 \hat P_2  + 1} \right)\]
\[   + \frac{{b^2 }}
{6}\left[ {\left( {\hat Q_1 \hat P_1 } \right)^2  + \left( {\hat
Q_2 \hat P_2 } \right)^2  - 5\left( {\hat Q_1 \hat P_1
 + \hat Q_2 \hat P_2 } \right) - 12\hat Q_1 \hat P_1 \hat Q_2 \hat P_2
   - 2} \right]   \]
\begin{equation}\label{3.3.13}
\begin{gathered}
   + c\left[ {\left( {\hat Q_1 \hat P_1 } \right)^2  +
   \left( {\hat Q_2 \hat P_2 } \right)^2  + 3\left( {\hat Q_1 \hat P_1
    + \hat Q_2 \hat P_2 } \right) + 4\hat Q_1 \hat P_1 \hat Q_2 \hat P_2
     + 2} \right] \hfill \\
\end{gathered}
\end{equation}
It is easy to see, that basis vectors $ \left| {NL} \right\rangle$
are eigenvectors for quantum normal form (\ref{3.3.13}). Therefore
we get the simple analytical formula for the energy spectrum in
the fourth approximation.
\begin{equation}\label{3.3.14}
E\left( {N,L} \right) = N + 1 + \frac{{b^2 }} {{12}}\left[ {7L^2 -
5\left( {N + 1} \right)^2  + 1} \right] + \frac{c} {2}\left[
{3\left( {N + 1} \right)^2  - L^2  + 1} \right]
\end{equation}
Assuming $c = 0$ in the formula (\ref{3.3.14}) we get approximate
spectrum of the Henon-Heiles Hamiltonian, which up to the constant
shift coincides with the spectra obtained by other methods. The
last property, apparently, is connected with the ambiguity of
quantization method of the Hamiltonian. Each level in the energy
spectrum (\ref{3.3.14}) is double degenerated with respect to sign
of angular momentum number $L$ , whereas for the levels of the
exact Hamiltonian with $L = 3k(k = 1,2,...)$ the degeneration must
be taken off. Inclusion in the normal form the members of higher
degree leads to the taking off the degeneration. Really
\begin{equation}\label{3.3.15}
 \begin{gathered}
  \Gamma ^{\left( 6 \right)}  = \left( {\frac{{11}}
{{54}}b^4  + \frac{{10}} {9}b^2 c + 2c^2 } \right)\left( {P_1^3
Q_1^3  + P_2^3 Q_2^3 }
\right) +  \hfill \\
   + \left( {\frac{5}
{{12}}b^4  - \frac{{61}} {3}b^2 c + 15c^2 } \right)P_1 P_2 Q_1 Q_2
\left( {P_1 Q_1
 + P_2 Q_2 } \right) +  \hfill \\
   + \frac{2}
{9}b^2 \left( {7b^2  - 26c} \right)\left( {P_1^3 Q_2^3
+ P_2^3 Q_1^3 } \right) \hfill \\
\end{gathered}
\end{equation}

The last term in this relation connects the states with equal $N$
and the states with $L' - L =  \pm 6$ . However in this
approximation the basis vectors $ \left| {NL} \right\rangle$
(\ref{3.3.8}) will be no longer the eigenvectors of the normal
form. Therefore for the calculation of the energy spectrum it is
required either additional diagonalization of the engaging states
or the calculations by the theory of perturbation \cite{63}. How
well does the energy spectrum (\ref{3.3.14}), obtained by
quantization of classical normal form in the fourth approximation,
reproduces exact quantum spectrum of the Hamiltonian of quadrupole
oscillations? In the Table 3.3.1 approximate spectrum $E(NL)$ is
compared with the spectrum $E_{ex} $ , obtained by diagonalization
on oscillator basis.
\begin{table}
\begin{center}
Table 3.3.1.  Energy levels of Hamiltonians (16.1)
($b=0.04416,c=0.00015, E_{cr}=90$)\\
\medskip
\begin{tabular}{|c|c|c|c|c|c|c|} \hline
&N&L&$E_{approx}$&$E_{exact}$&Type&Error $\Delta E/E$\%\\\hline
1.&0&0&0.9996&1.0001&$A1$&0.43\\\hline
2.&1&1&1.9989&1.9994&$E$&0.022\\\hline
3.&2&0&2.9949&2.9954&$A1$&0.015\\\hline
4.&2&2&2.9992&2.9996&$E$&0.014\\\hline
5.&3&1&3.9919&3.9924&$E$&0.012\\\hline
6.&3&3&4.0004&4.0008&$A1$&0,010\\\hline
&&&&4.0008&$A2$&0.010\\\hline
7.&4&0&4.9855&4.9861&$A1$&0.010\\\hline
8.&4&2&4.9898&4.9903&$E$&0.010\\\hline
9.&4&4&5.0025&5.0059&$E$&0.07\\\hline
10.&5&1&5.9801&5.9807&$E$&0.01\\\hline -&-&-&-&-&-&-\\\hline
145.&23&1&23.662&23.670&$E$&0.032\\\hline
146.&23&3&23.671&23.673&$A1$&0.009\\\hline
&&&&23.688&$A2$&0.071\\\hline
147&23&5&23.688&23.699&$E$&0.045\\\hline
148&23&7&23.713&23.722&$E$&0.037\\\hline
149&23&9&23.747&23.754&$A1$&0.028\\\hline
&&&&23.754&$A2$&0.028\\\hline
150&23&11&23.790&23.794&$E$&0.019\\\hline -&-&-&-&-&-&-\\\hline
246&30&10&30.541&30.558&$E$&0.055\\\hline
247&30&12&30.588&30.601&$A2$&0.043\\\hline
&&&&30.601&$A1$&0.043\\\hline
248&30&14&30.643&30.653&$E$&0.031\\\hline
249&30&16&30.707&30.712&$E$&0.017\\\hline
250&30&18&30.779&30.780&$A1$&0.003\\\hline
&&&&30.780&$A2$&0.003\\\hline
\end{tabular}
\end{center}
\end{table}
\begin{figure}
\centering
\includegraphics[height=12cm]{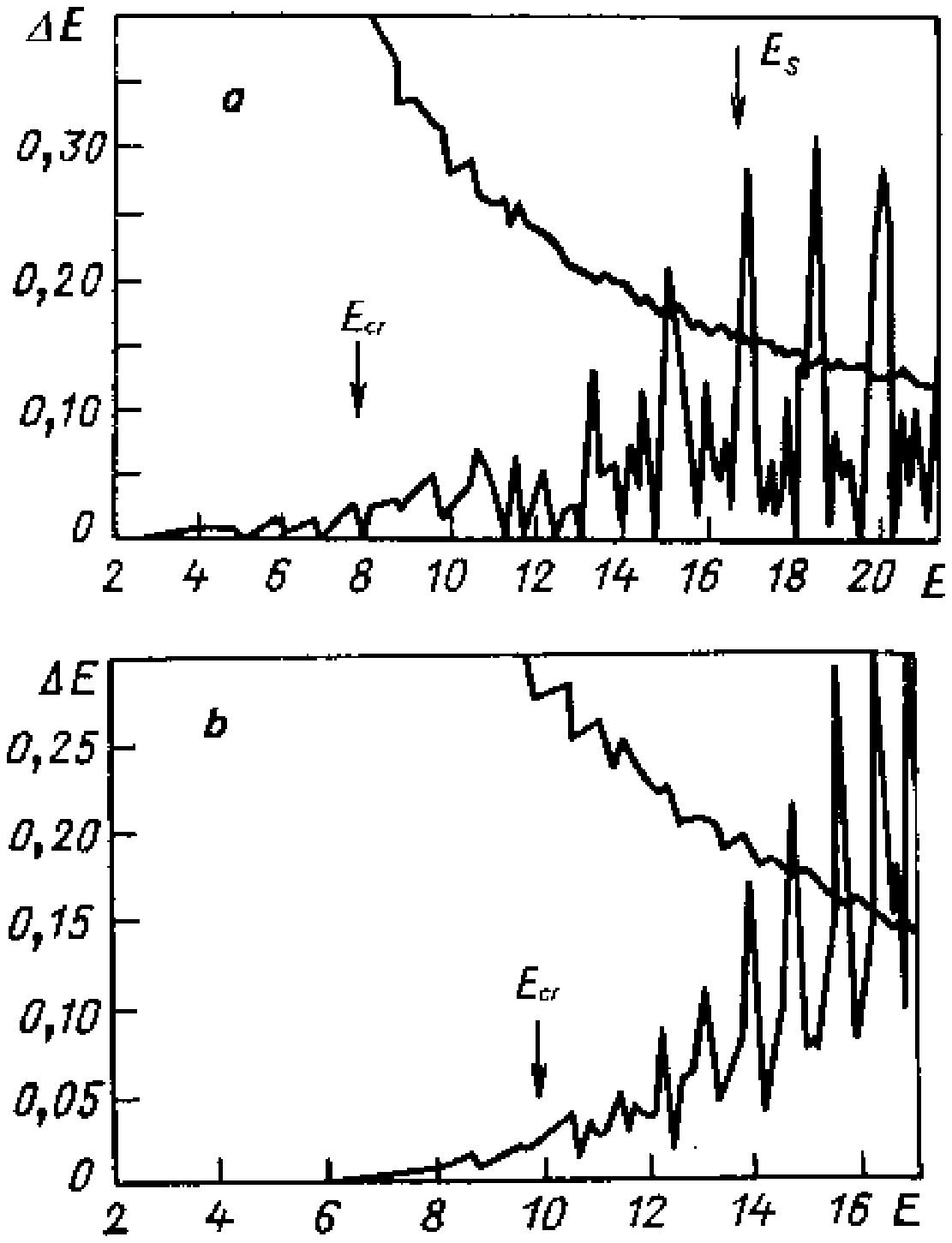}
\caption{Difference between exact quantum energy levels and
quasiclassical levels (\ref{3.3.14}). The line is the mean level
spacing. (a) for Henon-Heiles Hamiltonian with parameter $b = 0.1$
 $\left( {E_{cr}  = 8.3;\;E_S  = 16.7} \right)$
; (b) for Hamiltonian of quadrupole oscillations with parameters:
$W = 13;a = 1;c = 0.00135\left( {E_{cr}  = 10} \right)$.}
\label{Fig.20}
\end{figure}

The dimension of submatrices of type $A_1 ,A_2 ,E$ are
respectively equal $560,560$ and $415$ and provide sufficient
precision of calculations for the first $250 - 300$ levels. The
parameters of the Hamiltonian (16.1) are chosen so that $E_{cr1} =
90$ .  As it can be seen from the Table 3.3.1 in region, where
classical motion is regular, the quantum normal form reproduces
the energy spectrum with the relative error $\Delta E/E \sim
0.01\% $ . It seems naturally to expect that in the neighbourhood
of critical energy of the transition to chaos, where the
destruction of the approximate integrals of motion takes place and
with the help of which the semiclassical spectrum is built, the
agreement of the last one with the exact spectrum must essentially
be getting worse. Let us analyze this effect on the example of the
spectrum of quadrupole oscillations. The difference $ \Delta E =
\left| {E\left( M \right) - E_{ex} } \right|$ as the function of
energy calculated for the case $W = 13,E_{cr}  = 10$ is
represented in Fig.20(b). We can see that in the region of
energies, where the classical motion is regular, the approximate
spectrum (\ref{3.3.14}) reproduces the exact one rather well. At
the transition to chaotic region the difference increases sharply.
The similar situation takes place and for the Henon-Heiles
Hamiltonian (Fig.20(a)).

\subsection{$R - C - R$
 transition and statistical properties of energy spectrum of the
 QON}

Important correlations between peculiarities of classical dynamics
and structure of quantum energy spectrum can be obtained in
investigation of statistical properties of level sequences. We
shall be interested in local properties of spectrum, i.e. in
deviation in distribution of levels from mean values and in
fluctuations. Why should we address to local characteristics of
spectrum? The matter is, that the global characteristics like the
numbers of states $N(E)$ or the smoothed density of levels $\rho
(E)$ are too rough. At the same time, such local characteristics,
as the function of the nearest-neighbour spacing distribution
between levels is very sensitive to the properties of potential
and to the shape of boundary. It is sufficient, for example, to
bend slightly one of the walls in square to make it scattering,
then classical trajectories in such system become chaotic. In this
case the essential rearrangement of functions of the
nearest-neighbour spacing distribution between levels happens,
though the number of states $N(E)$ changes slightly or doesn't
change at all.

Relying on rather simple reasons \cite{65}, let's try to construct
one of the local characteristics of quantum spectrum - the
function of the nearest-neighbour spacing distribution between
levels $\rho (s)$ . For a random sequence the probability that a
level will be in the small interval $(E + s,E + s + ds)$ ,
proportional, of course, to $ds$ , will be independent of whether
or not there is a level at $E$ . This will be modified if we
introduce level interaction. Given a level at $E$ , let the
probability that the next level)$(s \ne 0)$ be in $(E + s,E + s +
ds)$ be $P(s)ds$ . Then for $P(s)$ , the nearest-neighbour spacing
distribution (NNSD), we have
\begin{equation}\label{3.4.1}
 P\left( S \right)dS = P\left( {\left. {1 \in dS} \right|0 \in S}
 \right)P\left( {0 \in S} \right)
\end{equation}
where $P(n \in S)$ is the probability that the interval of length
$S$ contains $n$ levels and the $P(\left. {n \in dS} \right|m \in
S)$ is conditional probability that the interval of length $dS$
contains $n$ levels, when that of length $S$ containes $m$ evels.
The second factor in (\ref{3.4.1}) is $ \int\limits_S^\infty
{P\left( x \right)dx}$ , the probability that the spacing is
larger than $S$ , while the first one will be $dS$ times a
function of $S,r_{10} (S)$ , depending explicitly on the choice,
$1$ and $0$ , of the discrete variables $n,m$ . Then
\begin{equation}\label{3.4.2}
P\left( S \right) = r_{10} \left( S \right)\int\limits_S^\infty
{P\left( x \right)dx}
\end{equation}
which we can solve easily to find
\begin{equation}\label{3.4.3}
P\left( S \right) = cr_{10} \exp \left( { - \int\limits_0^S
{r_{10} \left( x \right)dx} } \right)
\end{equation}
The Poisson law follows if we take $r_{10} (S) = 1/D$ (the absence
of correlations between level positions), where $D$ is the mean
local spacing so that $1/D$ is the density of levels. Wigner's law
follows from the assumption of a linear repulsion, defined by
$r_{10} (S) = \alpha S$ . The arbitrary constants are determined
by conditions
\begin{equation}\label{3.4.4}
\int {P\left( x \right)dx = 1,} \int {xP\left( x \right)dx = D}
\end{equation}
Finally, we find for the Poisson and Wigner cases, respectively
\begin{equation}\label{3.4.5}
P\left( S \right) = 1/D\exp \left( { - {\raise0.7ex\hbox{$S$}
\!\mathord{\left/
 {\vphantom {S D}}\right.\kern-\nulldelimiterspace}
\!\lower0.7ex\hbox{$D$}}} \right),\quad S \geqslant 0
\end{equation}
\begin{equation}\label{3.4.6}
P\left( S \right) = \frac{{\pi S}} {{2D^2 }}\exp \left( { -
\frac{{\pi S^2 }} {{4D^2 }}} \right),\quad S \geqslant 0\quad
\end{equation}
The second distribution displays the repulsion explicitly since
$P\left( 0 \right) = 0$ , in contrast to the Poisson form, which
has a maximum at $S = 0$ (level clusterization).

To get a first idea about origin of the level repulsion let us
consider \cite{65} the Hamiltonian as defined with respect to some
fixed basis by its matrix elements. The repulsion may be regarded
as arising from the fact that the subspace for which the
corresponding spectrum has a degeneracy is of a dimensionality
less by two than that of the general matrix-element space, so that
in some sense a degeneracy is "unlikely". Alternatively \cite{66},
if we think of the matrix elements as functions of a parameter
$\alpha $ , we cannot, in general, force a crossing by varying
$\alpha $ but we must instead take the matrix elements as
functions of at least two parameters which are independently
varied. In the one-parameter case one will find that, if two
levels approach each other as $\alpha $ is varied, then instead of
crossing they will turn away as if repelled.

There is a principal difficulty with the derivation of
(\ref{3.4.6}). Why should we assume a linear repulsion? Although
there are some plausibility arguments for this form, the result
cannot be correct for every system. Furthermore only a probability
arguments cannot explain the nature of level repulsion. However
the situation changes, if we intend that statistical properties of
the sequence of levels of real physical system are equivalent to
the sequences of eigenvalues of the ensemble of random matrices of
a definite symmetry. The theory, based on this hypothesis, was
completed in the sixties \cite{67}, The final result for the
function NNSD is
\begin{equation}\label{3.4.7}
 P\left( S \right) \approx S^\alpha  \exp \left( { - \beta S^2 } \right)
\end{equation}
The critical index $\alpha $ determining the behaviour of the
distribution function at $S \to 0$ depends on the symmetry of the
ensemble of matrices. This symmetry is determined by the
properties of the physical system, the spectrum statistics of
which we want to reproduce. If the system is invariant relative to
time reversion, then the corresponding ensemble is Gaussian
orthogonal ensemble (GOA). For the system assuming the violation
of the invariance relative to time reversion the Gaussian unitary
ensemble is associated. Finally, the symplectic ensemble of random
matrices corresponds to the Hamiltonian of more complicated
structure $ H = H_0  + \vec h\vec \tau $ , where $ H_0^ *   =
H_0^T ,h_K  = h_K^ *   =  - h_K^T $ , ($K = 1,2,3;\;\sigma _K $ -
Pauli matrices). The critical index $\alpha $ in the formula
(\ref{3.4.7}) is equal to: $\alpha  = 1$ for the GOA, $\alpha  =
2$ for the unitary ensembles and $\alpha  = 4$ for the symplectic
ensembles.

The predictions of the statistical theory of levels (especially
for the GOA) were compared in details with all available sets of
nuclear data \cite{65,67}. No essential deviations between the
theory and data base have been revealed. Similar comparisons have
been realized for atomic spectra. And here, a good agreement has
been revealed with the predictions of the GOA, though the number
of processed data was essentially less, than in the nuclear
spectroscopy.

If for the complicated systems (atomic nucleus, many-electron
atom) we can give serious arguments in favour of the hypothesis of
the equivalence of the statistical properties of the spectrum and
the sequence eigenvalues of the ensemble of the random matrices,
so its generalization does not seem natural in the case of the
systems with a small number of degrees of freedom.

A radically new and sufficiently universal approach to the problem
of the statistical properties of energy spectra may be developed
on the basis of a nonlinear theory of dynamic systems. The
numerical calculations \cite{23,68,69,70,71}, supported by
sophisticated theoretical considerations
\cite{1,4,6,23,54,56,72,73,74}, show that the important universal
peculiarity of the energy spectra of the systems, that are chaotic
in the classical limit, is the phenomenon of the level repulsion;
while the systems, whose dynamics is regular in the classical
limit, are characterized by the level clusterization. This
statement is sometimes called \cite{68} as the hypothesis of the
universal character of the fluctuations of energy spectra.

Among the systems, which spectra were subjected to detail
numerical analysis, the central place is occupied by
two-dimensional billiards (free particle moving on the plane
inside of some region and subjecting to elastic reflection on the
boundary). One of two extremal situations can be realized for the
billiards with the definite shape of the boundaries: integrable or
nonintegrable. The angular momentum is the second integral (except
energy) in the circular billiard and such a system is integrable.
The billiard like "stadium" is one of the simplest stochastic
systems \cite{110}.

The NNSD for the integrable system (the circular billiard) is well
approximated by Poisson distribution and the variance is the
linear function of the considered energy interval, that is in
complete correspondence with the hypothesis of the universal
character of the fluctuations of energy spectra. In the
nonintegrable case ("stadium"), the effect repulsion of levels and
slow growth of the variance, caused by the rigidity of
corresponding spectrum, are observed.

Measure of rigidity is the statistic $\Delta _3 $ of Dyson and
Mehta \cite{75}
\begin{equation}\label{3.4.8}
  \Delta _3 \left( {L;x} \right) = \frac{1}
{L}Min_{A,B} \int\limits_x^{x + L} {\left[ {n\left( \varepsilon
\right) - A\varepsilon  - B} \right]^2 dx}
\end{equation}
which determines the least-square deviation of the staircase
representing the cumulative density $n\left( \varepsilon  \right)$
from the best straight line fitting it in any interval $\left[
{x,x + L} \right]$ . The most perfectly rigid spectrum is the
picket fence with all spacing equal (for instance, the
one-dimensional harmonic oscillator spectrum),therefore maximally
correlated, for which $\Delta _3 (L) = 1/12$ , whereas, at the
opposite, the Poisson spectrum has a very  large average value of
$\Delta _3 \left( {\bar \Delta _3  = L/15} \right)$ , reflecting
strong fluctuations around the mean level density.

In contrast to billiards, where the character of motion does not
depend on energy, the Hamiltonian systems of general position are
the systems with the separable phase space, which contain both the
regions, where the motion is stochastic, and the islands of
stability. How is this circumstance reflected in statistical
properties of spectrum? Berry and Robnik \cite{73} and
independently Bogomolny \cite{76}, basing on the semiclassical
arguments, showed that NNSD for such system represents the
independent superposition of the Poisson distribution with the
relative weight $\mu $ , determining by the part of the phase
space with regular motion, and the Wigner distribution with the
relative weight $ \bar \mu \left( {\mu  + \bar \mu  = 1} \right)$
, determining by the part of the phase space with chaotic motion
\begin{equation}\label{3.4.9}
 P\left( x \right) = \mu ^2 \exp \left( { - \mu x} \right)erfc\left( {\frac{{\sqrt \pi  }}
{2}\bar \mu x} \right) + \left( {2\mu \bar \mu  + \frac{\pi }
{2}\bar \mu ^3 x} \right)\exp \left( { - \mu x - \frac{\pi }
{4}\bar \mu ^2 x^2 } \right)
\end{equation}
The expression (\ref{3.4.9}) represents the interpolated formula
between the Poisson (\ref{3.4.5}) and Wigner (\ref{3.4.6})
distributions.

Now let us go to the statistical properties of the spectrum of the
Hamiltonian of the quadrupole oscillations (\ref{3.2.1}). We
intend to study the evolution of these properties in the process
of $R - C - R$ transition, i.e. in each of energetic intervals
$R_1 ,\;C,\;R_2 $ . In this study we are restricted by the case of
one-well potentials ( $W < 16$ ). At the fixed topology of
potential surface ($W = const$ ) the unique free parameter of
quantum Hamiltonian (\ref{3.2.1}) is the scaled Planck's constant
$ \bar \hbar i $ . In the study of the concrete energetic interval
($R_1 ,C$ or $R_2 $ ), corresponding to definite type of classical
motion, the choice of $ \bar \hbar _i $ is dictated by the
possibility of attainment of the necessary statistical assurance
(the required number of levels in investigated interval) with
conservation of precision of spectrum calculation (restrictions to
possibility of diagonalization of matrices of large dimension).
Let us notice, that for the initial nonreduced Hamiltonian (16.1)
the variation of $ \bar \hbar _i $ is equivalent to the variation
of the parameters $a;b;c$ (i.e. variation of the critical energies
$E_{cr1} $ and $E_{cr2} $ ) or to the adequate choice of initial
Planck's constant $ \bar \hbar _i  \equiv \bar \hbar $ . The
values of the scaled Planck's constant $ \bar \hbar _3  \equiv\bar
\hbar $ , represented in Table 3.4.1, allow us to obtain in
corresponding energy intervals some hundreds of levels with the
precision better than $1\% $.
\begin{table}
\begin{center}
Table 3.4.1\\
\medskip
\begin{tabular}{|c|c|c|c|} \hline
   & $R_1 $ &$C$ & $R_2 $ \\ \hline
  $W = 13$ & $\bar E < 8 \cdot 10^{ - 5}$ & $8 \cdot 10^{ - 5}
    < \bar E < 8.4 \cdot 10^{ - 2}$ & $\bar E > 8.4 \cdot 10^{ - 2}$
  \\ \hline
 $ \bar \hbar $& $\bar \hbar  = 3.2 \cdot 10^{ - 6}$ & $\bar \hbar
 = 1.6 \cdot 10^{ - 4}$&$ \bar \hbar  = 1.1 \cdot 10^{ - 1}$ \\ \hline
\end{tabular}
\end{center}
\end{table}
\begin{center}
\end{center}
Two additional comments are necessary before proceeding to
results. We first recall that, to get rid of spurious effects of
the local properties due to variation of the density, one has to
work at constant density on the average. For this purpose one can
'unfold' the original spectrum \cite{65}, i.e. map the spectrum of
eigenvalues $\{ E_i \} $ onto the spectrum $\left\{ {\varepsilon
_i } \right\}$ through
\begin{equation}\label{3.4.10}
\varepsilon _i  = \bar N\left( {E_i } \right)
\end{equation}
Here the smoothed cumulative density $ \bar N\left( E \right)$ ,
\begin{equation}\label{3.4.11}
\bar N\left( E \right) = \int\limits_0^E {\bar \rho \left( {E'}
\right)dE'}
\end{equation}
and $ \bar \rho \left( E \right)$ is the smoothed level density.
In what follows we will take as energy unit the average spacing $
\bar x $ between two adjacent levels of the unfolded spectrum
\begin{equation}\label{3.4.12}
 \bar x = \bar s_i  = \left\langle {\left( {\varepsilon _{i + 1}
  - \varepsilon _i } \right)} \right\rangle
\end{equation}

Meaning of unfolding is understood by the following example
\cite{77}. There are a few spacings between states of the
same$(J^\pi  ,T)$ in the ground state region known for many
nuclei. Together they constitute a large sample of experimental
spacing data. The only trouble is that the underlying scaling
parameter $D$ varies from nucleus to nucleus. As a result, we
cannot make statistical studies unless we have a model for the
variation of $D$ as a function of nucleon number $A$ . This is a
kind of unfolding except, instead of energy, we must remove the
$A$ dependence of $D$ .

In the second place, we remark that our above discussion about
Poisson of Wigner level statistics applies only for a pure
sequence, i.e., one whose levels all have the same values of the
exact quantum number. More precisely, a pure sequence represents a
set of levels relating to one and the same nonreducible
representation of the group of symmetry of considered Hamiltonian.
In the case of mixed sequences \cite{65} the level repulsion is
moderated by the vanishing of Hamiltonian matrix elements
connecting two different symmetry states, the spectra have moved
towards random, and NNSD towards Poisson.
\begin{figure}
\centering
\includegraphics[height=14cm]{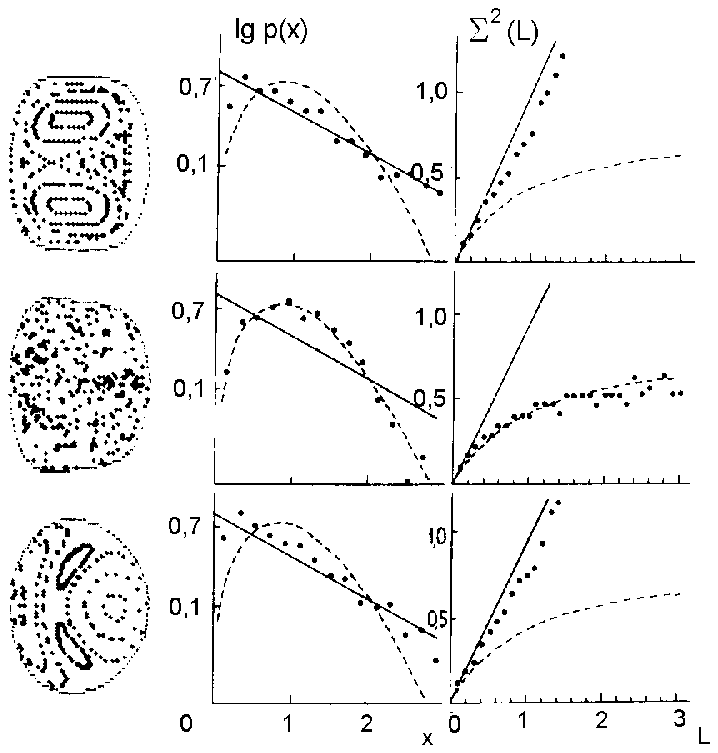}
\caption{Correlation between the character of classical motion and
statistical properties of energy spectra in $R - C - R$ transition
for the Hamiltonian of quadrupole oscillations. On the left: -
Poincare sections, in the middle - distribution function of the
nearest-neighbour spacings, on the right - the variance. Below -
the first regular range $\left( {R_1 } \right)$ , in the middle -
the chaotic range, above - the second regular range $\left( {R_2 }
\right)$ .} \label{Fig.21}
\end{figure}
The results of investigation of correlations between statistical
properties of energy spectrum of the Hamiltonian QON and the
character of classical motion are represented in Fig.21 \cite{78}.
The typical Poincare surfaces of section, showing the change of
classical regimes of motion, are represented at the left of
Figure, and the statistical characteristics of spectrum (logarithm
NNSD and dispersion), illustrating the evolution of the latter in
the process of the  $R - C - R$ transition, are represented at the
right. At the construction of statistical characteristics, a
purity of sequence is provided by using only those levels, which
are relative to definite nonreducible representation of $C_{3v} $
group (the levels of $E$ -type were used for results represented
in Fig.21; the statistical characteristics of levels of $A_1 $ and
$A_2 $ - have similar form).

Both the NNSD and the variance well correspond to the predictions
of GOE for the chaotic $(C)$ region . The logarithmic scale for
NNSD $p(x)$ is suitable to trace this correspondence at large $x$
. For regular regions $R_1 $ and $R_2 $ the distribution function,
in the same scale, according to the hypothesis of the universal
character of fluctuations of energy spectra, must be represented
by the straight line (the logarithm of Poisson's distribution).
The results demonstrate the agreement with this hypothesis, though
not large deviations are observed for small distances between
levels. Such a tendency to the rise of some repulsion in regular
region, apparently, is connected with a small admixture of chaotic
component.

The natural question about the role of numerical errors and, in
particular, about the role of basis cutting off arises in careful
studies of numerous examples confirming the hypothesis of the
universal character of fluctuations of energy spectra. Even the
special expression "the induced nonregularity" has appeared
\cite{79}. The investigation of statistical properties of spectrum
in three-levels model of Lipkin-Meshkov-Glick (LMG) \cite{80}
sheds a certain light on this problem.

Let us consider the system of $N$ particles on three levels, each
of which is degenerated with multiplicity $N$ . The Hamiltonian of
this system in representation of second quantization has the
following form
\begin{equation}\label{3.4.13}
H = \sum\limits_{k = 0}^2 {\varepsilon _k \left( {\sum\limits_{n =
1}^N {\alpha _{kn}^ +  \alpha _{kn} } } \right)}  - \frac{1}
{2}V\sum\limits_{k,l = 0}^n {\left( {1 - \delta _{kl} }
\right)\left( {\sum\limits_{n = 1}^N {\alpha _{kn}^ +  \alpha
_{\ln } } } \right)}
\end{equation}
where one-particle states (levels), marked by the indexes $k$ $(k
= 0,1,2)$ , and $n = 1,2,...N$ are different degenerate states
belonging to one level,$\alpha ^ +  ,\alpha $ are fermionic
creation and annihilation operators that obey the usual
anticommutation relations.

More traditional two-levels model LMG \cite{81} has unique degree
of freedom (it is the number of particles on top level) and
consequently its classical analog does not assume chaotic
behaviour. Introduction of an additional level is equivalent to
the appearance of additional degree of freedom. Stationary states
of the Hamiltonian (\ref{3.4.13}) can be marked by two indexes -
the numbers of particles on the second and the third levels. As
there is the finite number of ways of distribution of $N$
particles on three levels, then the number of basis vectors in
this model is finite. The problems of basis cutting off and
estimation of arising errors are absent.

Using coherent states
\begin{equation}\label{3.4.14}
\left| z \right\rangle  = \exp \left( {\sum\limits_{k = 1}^2
{\sum\limits_{n = 1}^N {z_k \alpha _{kn}^ +  \alpha _{0n} } } }
\right)\left| 0 \right\rangle
\end{equation}
where $ \left| 0 \right\rangle $ is the state with all $N$
particles on level with $k = 0$ , we determine the classical
Hamiltonian $H_{cl} $ as
\begin{equation}\label{3.4.15}
\begin{gathered}
  H_{cl}  \equiv \frac{1}
{{N\varepsilon }}\left\langle z \right|H\left| z \right\rangle  =
- 1 + \frac{1} {2}q_1^2 \left( {1 - \chi } \right) + \frac{1}
{2}q_2^2 \left( {2 - \chi } \right) + \frac{1} {2}p_1^2 \left( {1
+ \chi } \right) + \frac{1}
{2}p_2^2 \left( {2 + \chi } \right) +  \hfill \\
   + \frac{\chi }
{4}\left[ {\left( {q_1^2  + q_2^2 } \right)^2  - \left( {p_1^2 +
p_2^2 } \right)^2  - \left( {q_1^2  - p_1^2 } \right)\left(
{q_2^2  - p_2^2 } \right) - 4q_1 q_2 p_1 p_2 } \right] \hfill \\
\end{gathered}
\end{equation}
Here $\chi $ is the distance between levels, $\chi  = (N - 1)^V
/\varepsilon $ ,
\begin{equation}\label{3.4.16}
q_k  = \operatorname{Re} \frac{{\sqrt 2 z_k }} {{\sqrt {1 + \left|
{z_1 } \right|^2  + \left| {z_2 } \right|^2 } }},p_k  =
\operatorname{Im} \frac{{\sqrt 2 z_k }} {{\sqrt {1 + \left| {z_1 }
\right|^2  + \left| {z_2 } \right|^2 } }}
\end{equation}
The equations of motion for $q_k $ and $p_k $ , obtained from
variational principle \cite{7} are Hamiltonian. The classical
analog of the Hamiltonian LMG is integrable at $\chi  = 0\;(V =
0)$ . Changing the value of the constant of connection $\chi $ we
can observe the transition from regular motion to chaotic one. The
spectral fluctuations for three-levels model LMG given in Fig.22.
These results are the serious argument in favour of the hypothesis
of the universal character of fluctuations of energy spectra.
\begin{figure}
\centering
\includegraphics[height=14cm]{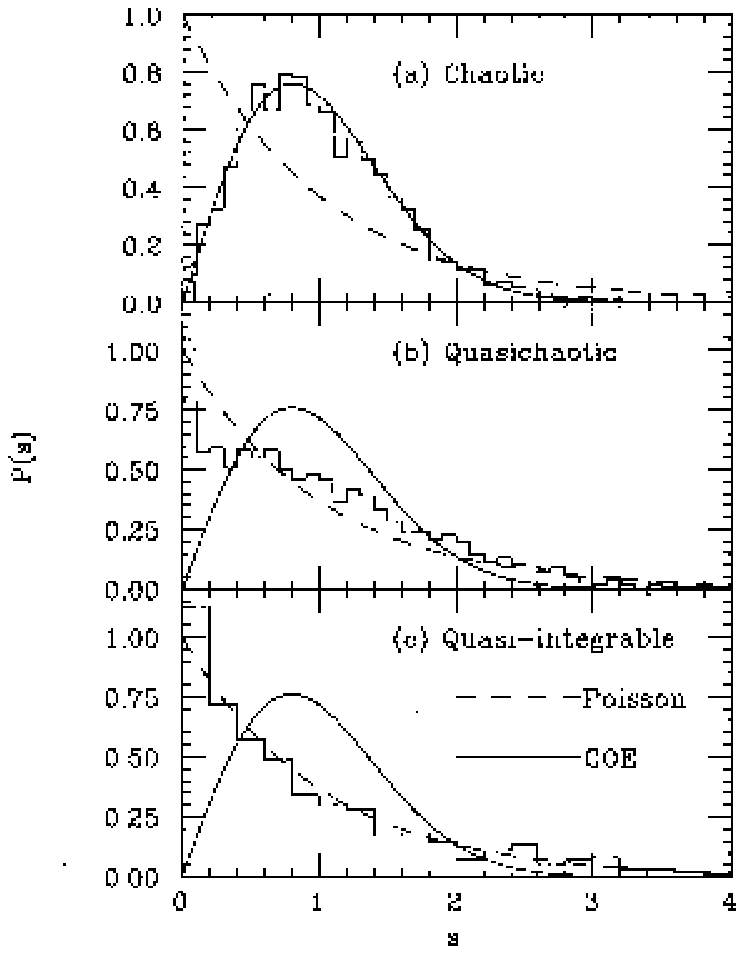}
\caption{The spectral fluctuations in three-levels model LMG: a -
distribution function of the nearest-neighbour spacings, b - the
variance [81].} \label{Fig.22}
\end{figure}

\subsection{  $R - C - R$
 transition and the structure of QON wave function}

As we have seen in the previous section, the statistical
properties of quantum spectrum of the Hamiltonian QON have been
rigidly correlated with the type of classical motion. It is
naturally to try to discover the analogous correlations in the
structure of wave functions, i.e. to assume, that the form of the
wave function for a semiclassical quantum state, associated with
classical regular motion in the regions $R_1 $ or $R_2 $ , is
different from the one for chaotic region $C$ . Notice that in
contrast to the description of the energy spectra where we now
have a sufficiently complete understanding of the spectral
statistical properties we are still far from a corresponding
complete knowledge of generic structure properties of the wave
functions. Furthermore it should be pointed out that in analysis
of QMCS at the level of energy spectra the principal role was
given to statistical characteristics, i.e., the quantum chaos was
treated as a property of the group states. The choice of a
stationary wave function of the quantum system, which is chaotic
in the classical limit, as an basic object of investigation
relates the phenomenon of quantum chaos to an individual state.

In contrast to spectrum the form of wave functions depends on the
basis in what they are determined. Studying QMCS the three
following representations are used more often

1. The so called $H_0 $ -representation is: the representation of
eigenfunctions $\left\{ {\varphi _n } \right\}$ of integrable part
$H_0 $ of total Hamiltonian $H = H_0  + V$ . The main objects of
investigation in this case are the coefficients of expansion
$C_{mn} $ of stationary functions $\psi _m $ in basis $\left\{
{\varphi _n } \right\}$ . $H_0 $ -representation is natural at
realization of numerical calculations, as diagonalization of
Hamiltonian $H$ is realized more often just in this
representation.

2. Coordinate representation in which the behaviour of wave
functions simply allows visual comparison with the picture of
classical trajectories in coordinate space.

3. Representation with the help of Wigner's functions \cite{82}
has a set of properties, common with classical function of
distribution in phase space.

As early as 1977 Berry \cite{83} assumed, that the form of the
wave function $\psi $ for a semiclassical regular quantum state
(associated with classical motion on a $N$ -dimensional torus in
the $2N$ -dimensional phase space) was very different from the
form of $\psi $ for an irregular state (associated with stochastic
classical motion on all or part of the $\left( {2N - 1} \right)$
-dimensional energy surface in phase space). For the regular wave
functions the average probability density in the configuration
space was seen to be determined by the projection of the
corresponding quantized invariant torus onto the configuration
space, which implies the global order. The local structure is
implied by the fact that the wave function is locally a
superposition of a finite number of plane waves with the same wave
number as determined by the classical momentum. In the opposite
case for the chaotic wave functions the averaged (over small
intervals of energy and coordinates) square of the eigenfunctions
in the semiclassical limit $ \hbar  \to 0 $ coincides with the
projection of the classical microcanonical distribution to the
coordinate space. Its local structure is spanned by the
superposition of infinitely many plane waves with random phases
and equal wave numbers. The random phases might be justified by
the classical ergodicity and this assumption immediately predicts
locally the Gaussian randomness for the probability of amplitude
distribution. Such structure of wave function is in well agreement
with the picture of classical phase space: the classical
trajectory homogeneously fill isoenergetic surface in condition
corresponding to chaotic motion. By contrast, from the
consideration of regular quantum state as analog of classical
motion on torus, a conclusion should be done about the singularity
(in limit $ \hbar  \to 0 $ ) of wave function near caustics
(boundaries of region of classical motion in coordinate space).

Berry's hypothesis was subjected to the most complete test for
billiards of different types and, in particular, for billiard
stadium type \cite{84}. The amplitude of typical wave function of
integrable circular billiard is negligible in classical prohibited
region (from conservation of angular moment in circular billiard
it should be that the arbitrary trajectory is enclosed between
external and certain internal circle, the radius of which is
determined by the angular moment) , whereas near the caustics it
is maximal. Distribution $ \left| \psi  \right|^2 $ for the case
stadium, at which classical dynamics is stochastic, is strongly
distinguished from the integrable case. However, this distribution
is not so homogeneous, as we could expect starting from the
ergodicity of classical motion.

Investigating QMCS in connection with properties of wave functions
let us pass from billiards to Hamiltonian system of general
position \cite{85,86}. For this purpose let us return to the
object of our investigation: $C_{3v} $ symmetric Hamiltonian QON.
We begin to study the correlations of our interest by the
topography of level curves of the stationary wave functions and,
in particular, of nodal curves on which $\psi _k (x,y) = 0$ . One
of a nonrigorous criterion of stochasticity \cite{87} states that
the system of nodal curves of regular wave function is a lattice
of quasiortogonal curves or is similar to such a lattice. At the
same time, the wave functions of chaotic states do not have such a
representation. Fig.23 confirms that the structure of the lattice
of nodal curves of separable wave functions undergoes a change in
the $R - C - R$ transition. The spatial structure of nodal curves
for states from regions $R_1 $ and $R_2 $ of regular classical
motion is considerably simpler than this structure for states from
region $C$ of chaos.
\begin{figure}
\centering
\includegraphics[height=14cm]{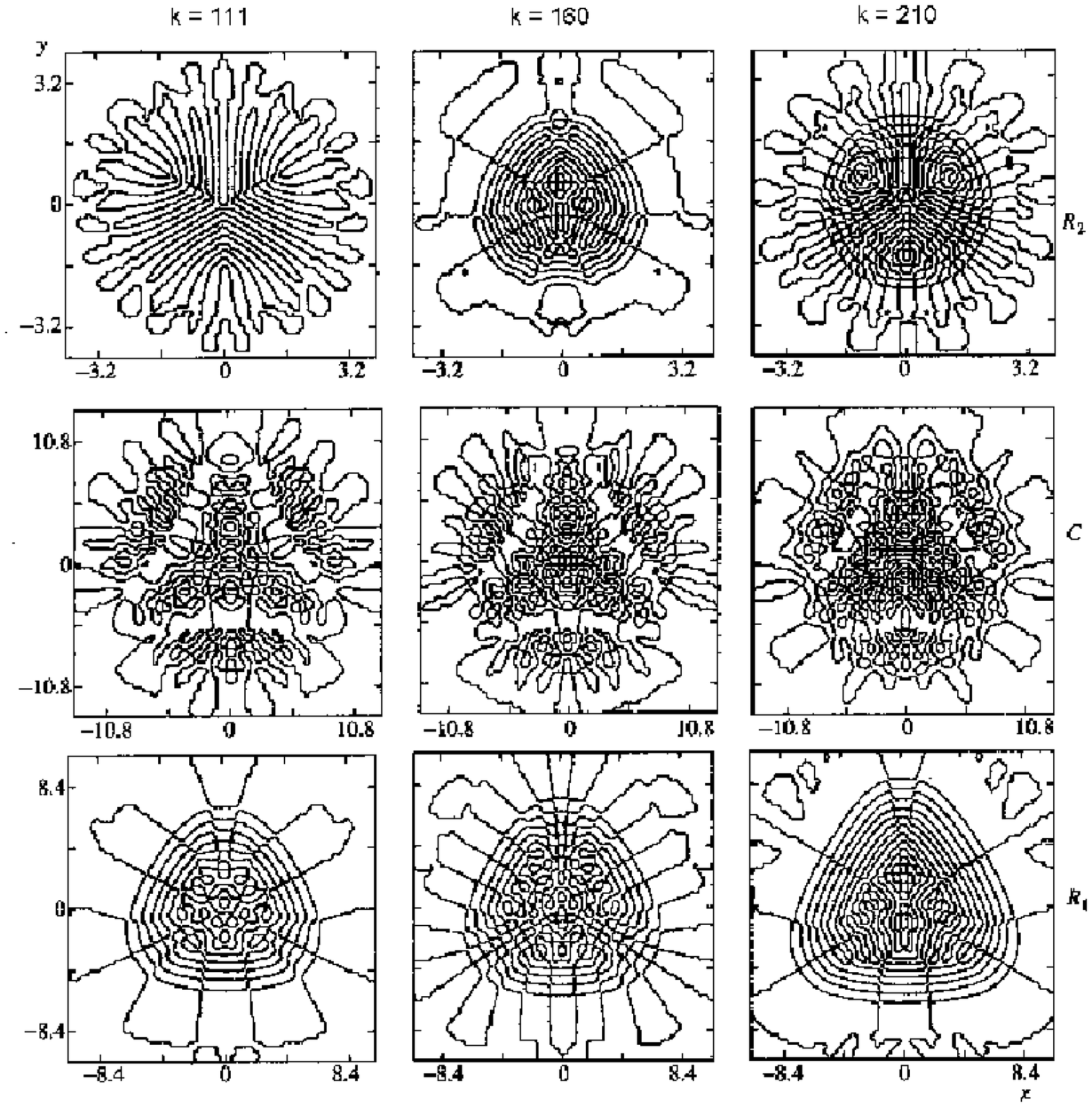}
\caption{Nodal curves of the wave functions $\psi _k \left( {x,y}
\right)$ . The numbers $k$ of the corresponding levels are shown
over the figures.} \label{Fig.23}
\end{figure}

Correlations between the structure of wave function and the type
of classical motion are also demonstrated in Fig.24, in which the
probability density $ \left| {\psi \left( {x,y} \right)} \right|^2
$ for states with numbers $111,160$ and $210$ is represented. The
squared module of the wave functions reproduces rather well
transition from functions with clear internal structure (region
$R_1 $ ) to an irregular distribution (region $C$ ) and the
restoration of structure in the second regular region ($R_2 $ ).
For the chosen technique, in which transition is traced for the
wave function with fixed number (because of our use of the scaled
Planck constant), a change in the wave function is associated only
with $R - C - R$ transition.
\begin{figure}
\centering
\includegraphics[height=14cm]{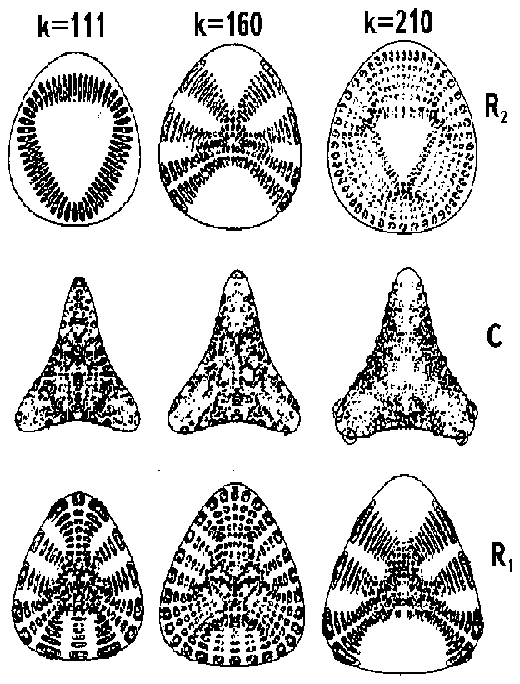}
\caption{Isolines of probability density $ \left| {\psi _k \left(
{x,y} \right)} \right|^2 $ . The step between isolines is
constant. The numbers $k$ of the corresponding levels are shown
over the figures.} \label{Fig.24}
\end{figure}

Evolution of the wave functions in the process of $ R - C - R $
transition can be studied also in $H_0 $ -representation or more
exactly in the representation of linear combinations of wave
functions of two-dimensional harmonic oscillator with equal
frequencies (see Section 3.2)
\begin{equation}\label{3.5.1}
\Psi _k  = \sum\limits_{N,L} {C_{NLj}^k \left| {NLj} \right\rangle
}
\end{equation}

If one introduces the notion of distributivity of wave function in
basis, then the criterion of stochasticity formulated by Nordholm
and Rise \cite{88} even in 1974 states that with the degree of
stochastization in the average arises the degree of distributivity
of wave functions. It is clear that this criterion is a direct
analog of the Berry's hypothesis for $H_0 $ - representation, if
one interprets the number of basis state $i = \left\{ {NLj}
\right\}$ as a discrete coordinate. Fig.25 qualitatively confirm
validity of this criterion. It can be seen from this figure that
the states that correspond to regular motion are distributed in a
relatively small number of basis states. At the same time, states
corresponding to chaotic motion (region $C$ ) are distributed in a
considerably larger number of basis state. In the latter case, the
contributions from a large number of basis states in expansion
(\ref{3.5.1}) interfere; this results in a complex spatial
structure of the wave function  $\psi _k \left( {x,y} \right)$ .
\begin{figure}
\centering
\includegraphics[height=8cm]{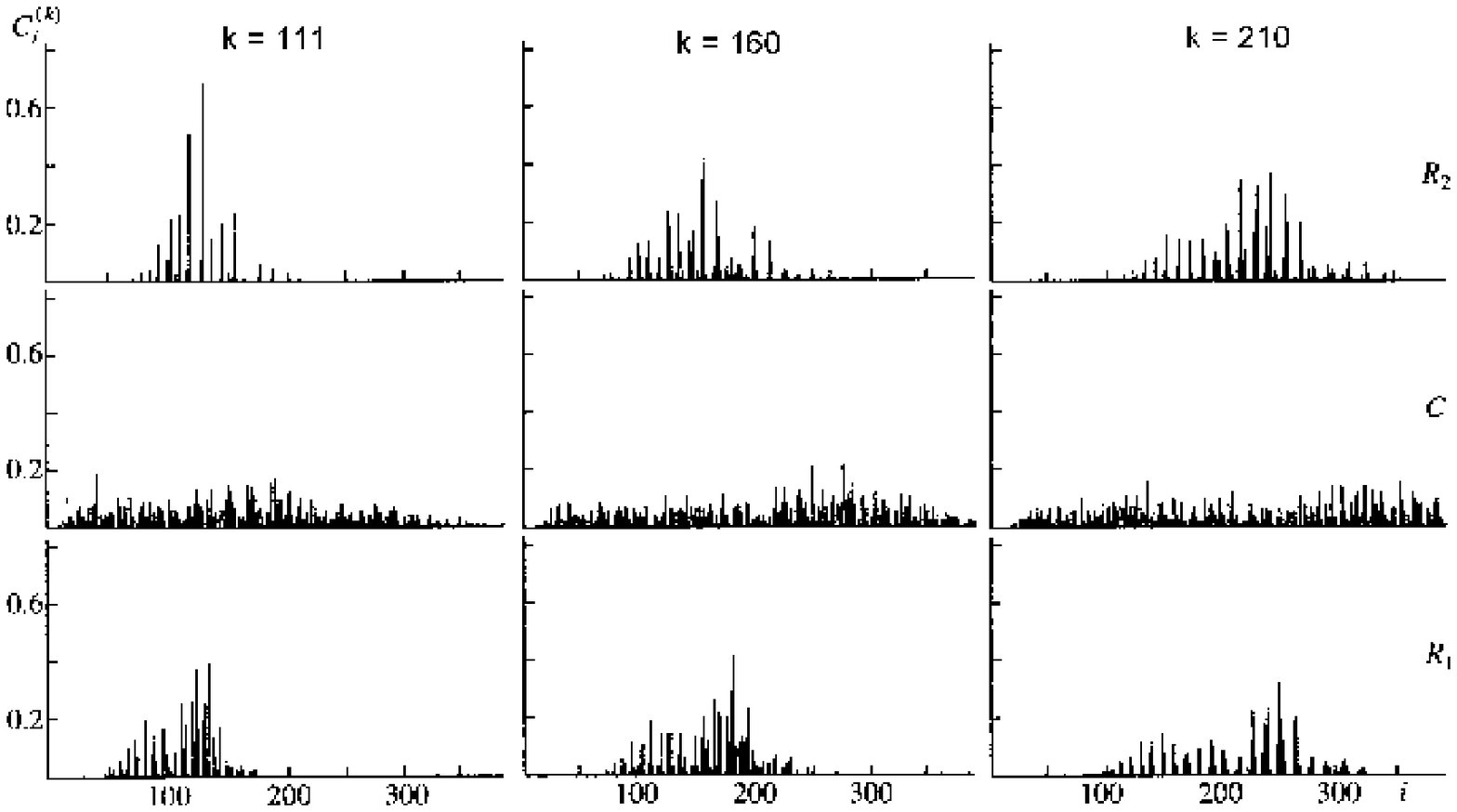}
\caption{Distribution of the coefficient $C_i^{\left( k \right)} $
in the number $i = \left\{ {N,L,j} \right\}$ of the basis state.
The numbers $k$ of the corresponding levels are shown over the
figures.} \label{Fig.25}
\end{figure}
For quantitative estimation of a degree of distributivity of wave
functions, it is useful to introduce \cite{89,90} the analog of
usual thermodynamic entropy
\begin{equation}\label{3.5.2}
S^k  =  - \sum {\left| {C_{NLj}^k } \right|^2 \ln \left|
{C_{NLj}^k } \right|^2 }
\end{equation}
\begin{figure}
\centering
\includegraphics[height=16cm]{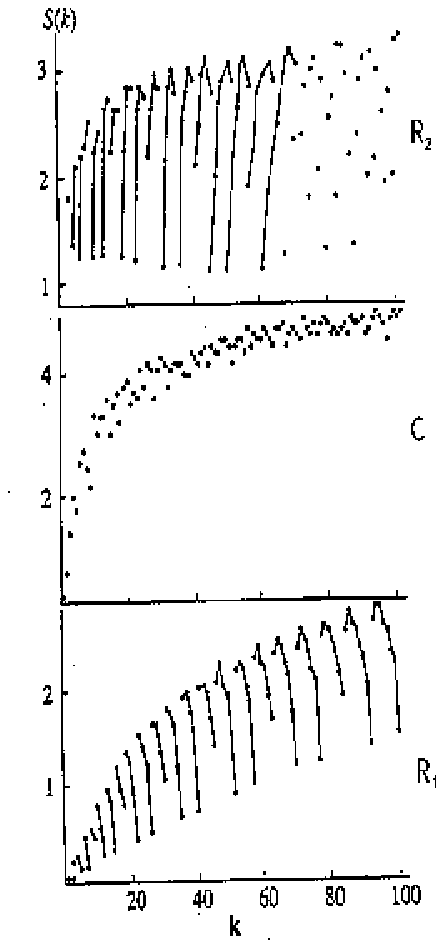}
\caption{Entropy $S$ as a function of the state number. The
stright lines connecting the points correspond to shell
classification according to $N$ . } \label{Fig.26}
\end{figure}
Fig.26 represents the dependence of entropy $S$ from the number of
state k for the first regular $\left( {R_1 } \right)$ and chaotic
$\left( C \right)$ regions. From this figure we can see, that the
type of classical motion in large measure determines the character
of this dependence: in chaotic region the entropy is practically
constant, that corresponds to the high degree of distributivity,
and in regular region the entropy is less in the value and has
nonmonotone character. Nonmonotonicity of $S$ is connected with
rather high degree of localization of the wave function depending
on the number of state. This effect and also the behaviour of the
entropy in the second regular region $R_2 $ , will be studied in
detail in the following section.

\subsection{ Evolution of shell structure in the process of  $R
- C - R$
 transition}

In sections 3.4 and 3.5 we have studied the change of statistical
roperties of energy spectra and structure of wave functions in the
process of $R - C - R$ transition. However, the utility of the
concept of the stochasticity extends further. Introduction of this
concept in the nuclear theory made possible to take a fresh look
\cite{13} at the old paradox \cite{22}: how one could reconcile
the liquid drop - short mean free path - model of the nucleus with
the independent particles - gas - like shell model. For the
solution of this paradox within the limits of philosophy of simple
chaos (see section 1.1(part 1)) it is sufficient to assume
\cite{13}

1) When the nucleonic motion inside the nucleus is integrable, one
expects to see strong shell effects in nuclear structure, quite
well reproducible, for example, by the model of independent
particles in the potential well.

2) When nucleonic motion is chaotic, one expects smooth,
statistical, Thomas-Fermi, Droplet model approaches to be good
approximations. This is because phase space is much more uniformly
filled with the Boltzmann fog in this case. At such approach the
elucidation of the mechanism of destruction of shell effects in
the process of the transition regularity-chaos plays the key role.
More appropriately the problem can be formulated in the following
way \cite{13}: how do shell dissolve with deviations from
integrability or, conversely, how do incipient shell effects
emerge as the system first begins to feel its proximity to an
integrable situation?

As it has been mentioned many times, the finite motion of
integrable Hamiltonian system with $N$ -degrees of freedom is, in
general, conditionally periodic, and the phase trajectories lie on
$N$ -dimensional tori. In the variables action-angle $\left(
{I,\theta } \right)$ the Hamiltonian is cyclic with respect to
angle variables, $H = H_0 \left( I \right)$ . Even at the
beginning of the century, Poincare called the main problem of
dynamics the problem about the perturbation of conditionally
periodic motion in the system defined by the Hamiltonian
\begin{equation}\label{3.6.1}
H = H_0 \left( I \right) + \varepsilon V\left( {I,\theta } \right)
\end{equation}
where $\varepsilon $ is a small parameter. The essential step in
the solution of this problem was the KAM theorem, asserting, that
at switching on nonintegrable perturbation, the majority of
nonresonance tori, i.e. the tori for which
\begin{equation}\label{3.6.2}
\sum\limits_{i = 1}^N {n_i \frac{{\partial H_0 }} {{\partial I_k
}} \ne 0}
\end{equation}
is conserved, distinguishing from the nonperturbed cases only by
small (to the extent of smallness of $\varepsilon $ ) deformation.
As we have seen in section 2.3(part 1), at definite conditions the
KAM formalism allows to remove out of Hamiltonian the members
depending on angle, using the convergent sequence of canonical
transformations \cite{3}. When it is succeed, we immediately find
that perturbed motion lies on rather deformed tori so that
trajectories, generated by perturbed Hamiltonian, remain
quasiperiodic. In other words the KAM theorem reflects an
important peculiarity of classical integrable systems to conserve
regular behaviour even at rather strong nonintegrable
perturbation. In the problem of our interest, concerning the
destruction of shell structure of quantum spectrum, the KAM
theorem also will be able to play an important role. Considering
the residual nucleon-nucleon interaction as nonintegrable addition
to selfconsistent field, obtained, for example, in Hartree-Fock
approximation, one can try to connect the destruction of shell
with the deviation from integrability. The existence of shell
structure at rather strong residual interaction (or at large
deformation) can be obligated to rigidity of KAM tori,
contributing to survival of regular behaviour. Such assumption
seems rather natural, especially if one takes into account, that
the procedure of quasiclassical quantization itself \cite{91,92}
is indirectly based on convergence of the same sequence of
canonical transformations, as the KAM theorem.

The aim of this section is to trace the evolution of the structure
of QON Hamiltonian. In numerical calculations of this section it
will be more convenient for us to use nonscale version of the
Hamiltonian (16.1). By nonperturbed Hamiltonian $H_0 $ we will
mean the Hamiltonian of two-dimensional harmonic oscillator with
equal frequencies: $H_0  \equiv H\left( {a = 1,b = 0,c = 0}
\right)$ . Its degenerate equidistant spectrum is well known. At
switching on perturbation, the degeneration disappears and the
shell structure forms. The number of states, for example, the
states of $E$ - type (the numerical results, represented below,
are relative to the states precisely of this type, though
analogous results take place for the states belonging to another
nonreducible representations of $C_{3v} $ - group) for the given
quantum number $N$ is equal to $ \frac{1} {2}\left( {N_1  +
Mod\left( {N,2} \right)} \right)$ , where $ N_1  = \frac{1}
{3}\left( {2N + Mod\left( {N,3} \right)} \right)$ , and $Mod\left(
{N,M} \right)$ is a remainder from division of $N$ by $M$ .

It is obvious, that eigenfunctions of exact Hamiltonian QON are
not the eigenfunctions of operators $\hat N$ and $\hat L$ already.
Nevertheless, as numerical calculations show even at rather large
nonlinearity, one can use the quantum number $N$ and $L$ for the
classification of wave functions. The measure of nonlinearity, at
which such classification (i.e. the existence of shell structure)
remains reasonable, is connected with quasicrossing of
neighbouring levels. Under the quasicrossing we shall understand
the approach of levels up to the distances of degree of precision
of numerical calculations (the detailed analysis of quasicrossing
see lower).
\begin{figure}[t]
\centering
\includegraphics[height=8cm]{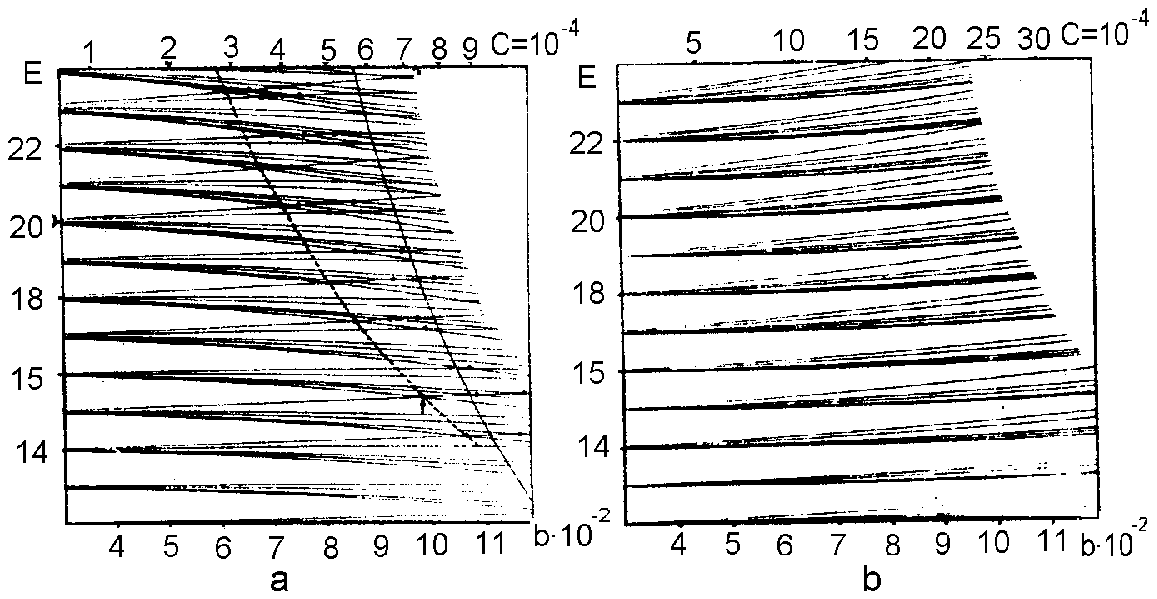}
\caption{a. Energy spectra of Hamiltonian (16.1) depending on
values of parameter $b$ for region $C$ $\left( {W = 13} \right)$ .
Points mark quasicrossinds. Continuous line - dependence $E_{cr} $
from $b$ . Dashed line shows the beginning of the region of
quasicrossings. Arrow shows point of quasicrossing of the levels
with $k = 40$ and $41$ . b. Energy spectra of Hamiltonian (16.1)
depending on values of parameter $b$ for region $R_1 \left( {W =
3.9} \right)$ . There are no quasicrossing of the levels. }
\label{Fig.27}
\end{figure}

The dependence of energy spectra of Hamiltonian QON from the
parameter $b$ for the values $W = 3.9$ and $W = 13$ are
represented in Fig.27a,b. As it is seen from Fig. 27a. for the PES
with $W = 13$ at the approaching to the line of critical energy of
the transition to chaos, defined according to the criterion of
negative curvature, the destruction of shell structure, which we
understand as multiple beginning of quasicrossing, takes place. At
the same time, for the PES with $W = 3.9$ in Fig.27b (for which as
we have shown in Section 2.4(part 1), the local instability is
absent and the motion is regular at all energies) the
quasicrossings are absent even at larger nonlinearity than in
Fig.27a for $W = 13$ .
\begin{figure}[t]
\centering
\includegraphics[height=10cm]{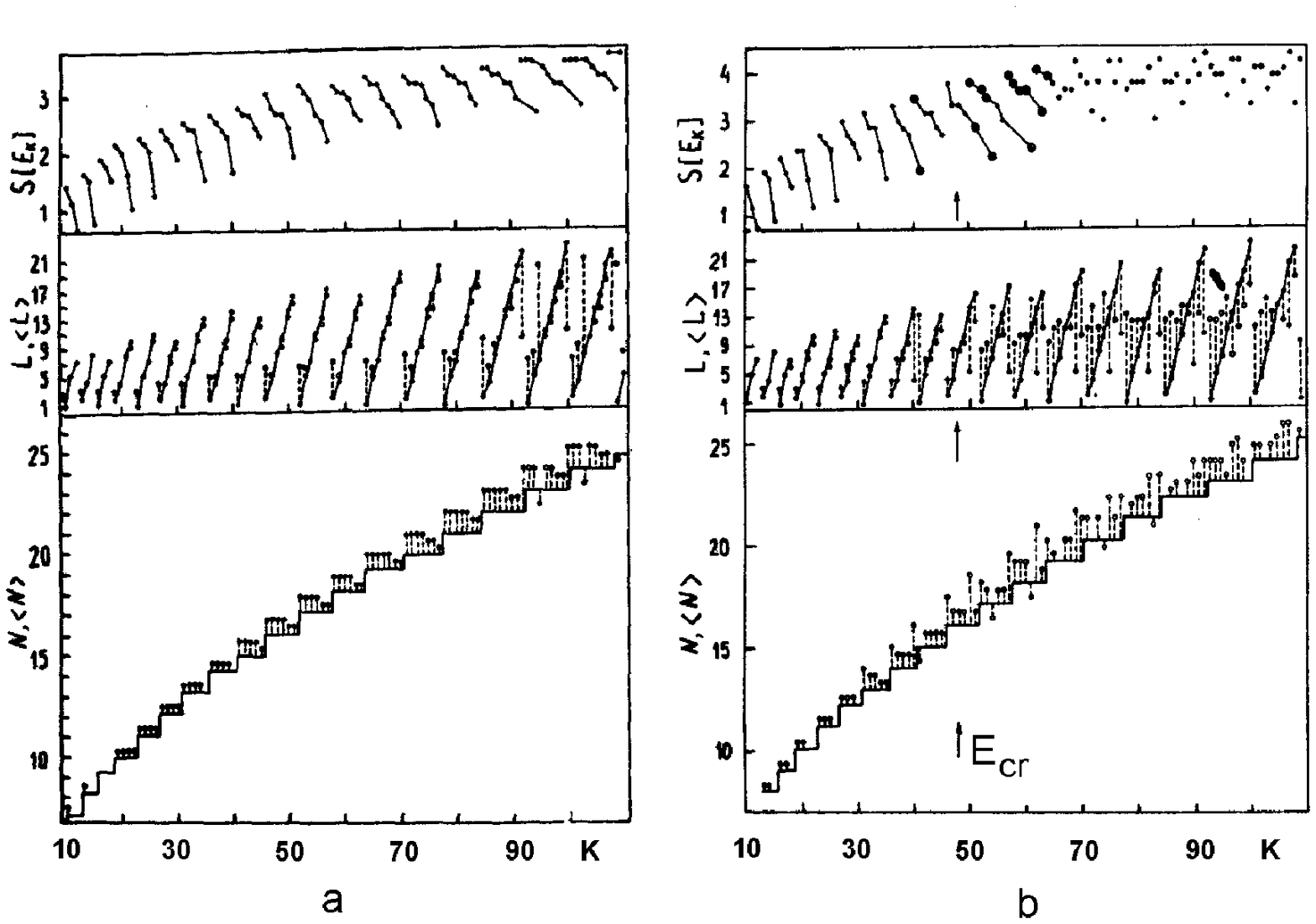}
\caption{Entropy $S$ and quantum numbers $L$,$N$ as the functions
of state number $k$.} \label{Fig.28}
\end{figure}

The destruction of shell structure can be traced for the wave
functions, using introduced in previous section, the analog of
thermodynamic entropy $S\left( {E_k } \right)$ . The character of
changes of entropy in regions $R_1 $ and $R_2 $ , corresponding to
regular classical motion, correlates with the transition from
shell to shell (see Fig.28). Two effects are observed in the
region $C$ , corresponding to chaotic classical motion. Firstly,
quasiperiodical dependence of entropy from energy is violated,
that testifies to destruction of sell structure. Secondly, the
monotone growth of entropy is observed on average with going out
on plateau corresponding to the entropy of random sequence at the
energies essentially exceeding critical energy.

Average values of operator $ \hat N$ , given in Fig. 29, and
calculated on stationary wave functions of exact Hamiltonian,
describe the dynamics of change of the shell structure. From this
Figure we can see, that the minimum deviation of $\left\langle N
\right\rangle $ from $N$ is observed in regions $R_1 $ and $R_2 $
, while this deviation is considerably greater in the stochastic
region $C$ ; moreover in the average it monotonically increases
with increase of energy. It is obvious, that at energies, for
which $N - \left\langle N \right\rangle  \geqslant 1$ , the
destruction of quantum analogs of classical  integrals leads to
the fact that the classification of levels with the help of
quantum numbers $N$ and $L$ loses sense.
\begin{figure}
\centering
\includegraphics[height=14cm]{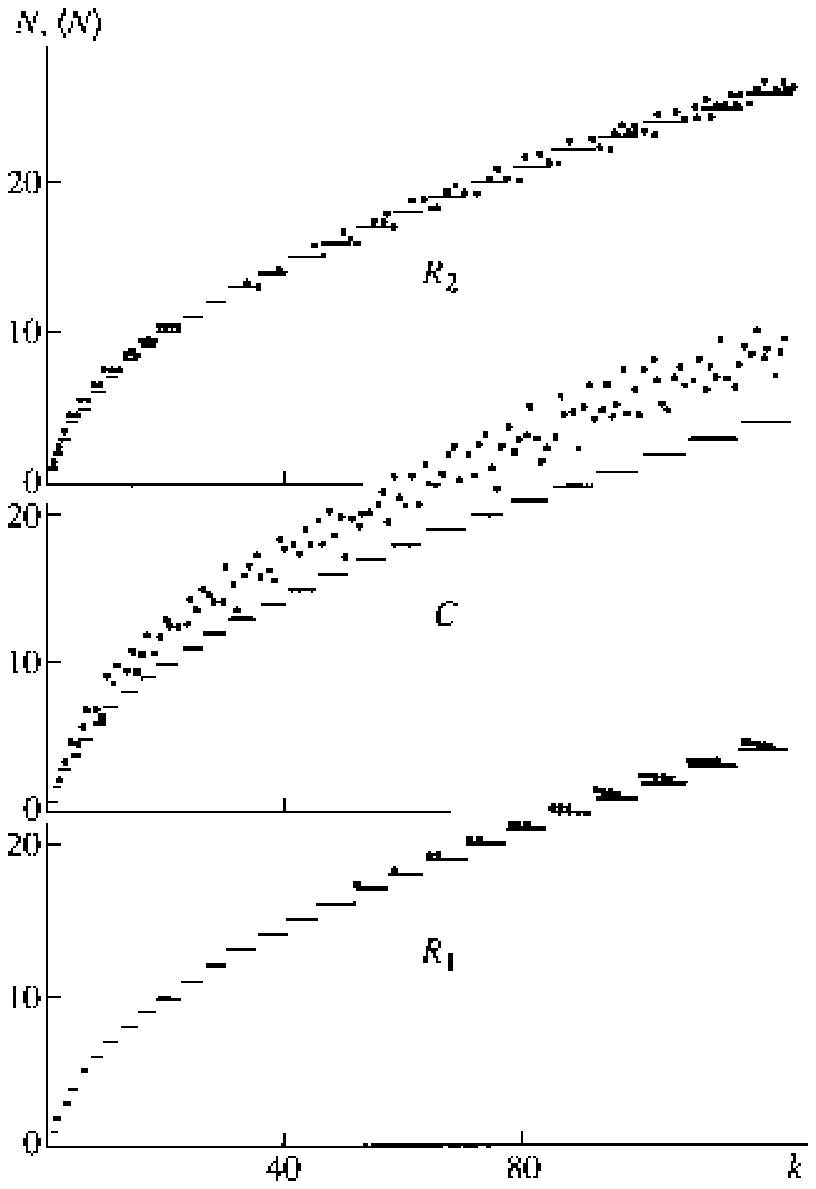}
\caption{Dependence of $N$ (lines) and $\left\langle N
\right\rangle $ points) of state numbers $k$ . } \label{Fig.29}
\end{figure}

Notice, that we could observe restoration of shell structure at
high energies in the process of $R - C$ transition only due to the
optimization of basis frequency (see Section 1.2), that is
equivalent to the possibility of diagonalization of matrices of
considerably higher dimension.

Now let us return to the analysis of quasicrossings of levels
arising near critical energy of the transition to chaos. As it is
known \cite{93}, for the set of Hamiltonians depending on two
parameters $\lambda  = \left( {b,c} \right)$ and that is invariant
with respect to inversion of time in the neighbourhood of
degeneration point $\lambda ^ *   = \left( {b^ *  ,c^ *  }
\right)$ , the difference of two terms $\Delta E = E_1  - E_2 $ is
determined by the expression
\begin{equation}\label{3.6.3}
\Delta E\left( {b,c} \right) = \left[ {A\left( {b - b*} \right)^2
+ B\left( {b - b*} \right)\left( {c - c*} \right) + C\left( {c -
c*} \right)^2 } \right]^{{\raise0.7ex\hbox{$1$} \!\mathord{\left/
 {\vphantom {1 2}}\right.\kern-\nulldelimiterspace}
\!\lower0.7ex\hbox{$2$}}}
\end{equation}
where coefficients $A,B,C$ are the functions of components
$\nabla_\lambda  H $ in the point $\lambda ^ *  $ . The
characteristics of nuclear shape \cite{91} in the problems of
nuclear spectroscopy more often play the role of parameters. The
wave functions of intersecting levels near degeneration point can
be represented in the following form
\begin{equation}\label{3.6.4}
\begin{gathered}
  \left| {E_1 \left( \lambda  \right)} \right\rangle  = \cos \chi _1 \left| {E_1 \left( {\lambda ^ *  } \right)} \right\rangle  + \sin \chi _1 \left| {E_2 \left( {\lambda ^ *  } \right)} \right\rangle  \hfill \\
  \left| {E_2 \left( \lambda  \right)} \right\rangle  = \cos \chi _2 \left| {E_1 \left( {\lambda ^ *  } \right)} \right\rangle  + \sin \chi _2 \left| {E_2 \left( {\lambda ^ *  } \right)} \right\rangle  \hfill \\
\end{gathered}
\end{equation}
where angles of mixing $\chi _{1,2} $ are determined by the
relation
\begin{equation}\label{3.6.5}
\begin{gathered}
  e^{2i\chi _{1,2} }  =  \pm \frac{{\left( {T\lambda } \right) + i\left( {S\lambda } \right)}}
{{\left[ {\left( {T\lambda } \right)^2  + \left( {S\lambda }
\right)^2 } \right]^{{\raise0.7ex\hbox{$1$} \!\mathord{\left/
 {\vphantom {1 2}}\right.\kern-\nulldelimiterspace}
\!\lower0.7ex\hbox{$2$}}} }}, \hfill \\
  \left( {T\lambda } \right) \approx H_{11} \left( \lambda  \right) - H_{22} \left( \lambda  \right),\left( {S\lambda } \right) \approx 2H_{21} \left( \lambda  \right) \hfill \\
\end{gathered}
\end{equation}
It is easy to see, that changing $ \lambda  \to  - \lambda $ the
"exchange" of wave functions of approaching levels takes place $
\left| {E_1 \left( {\lambda *} \right)} \right\rangle
\leftrightarrow \left| {E_2 \left( {\lambda *}
\right)}\right\rangle $ . Such test can be effectively used for
the analysis of nature of quasicrossing points.
\begin{figure}[t]
\centering
\includegraphics[height=10cm]{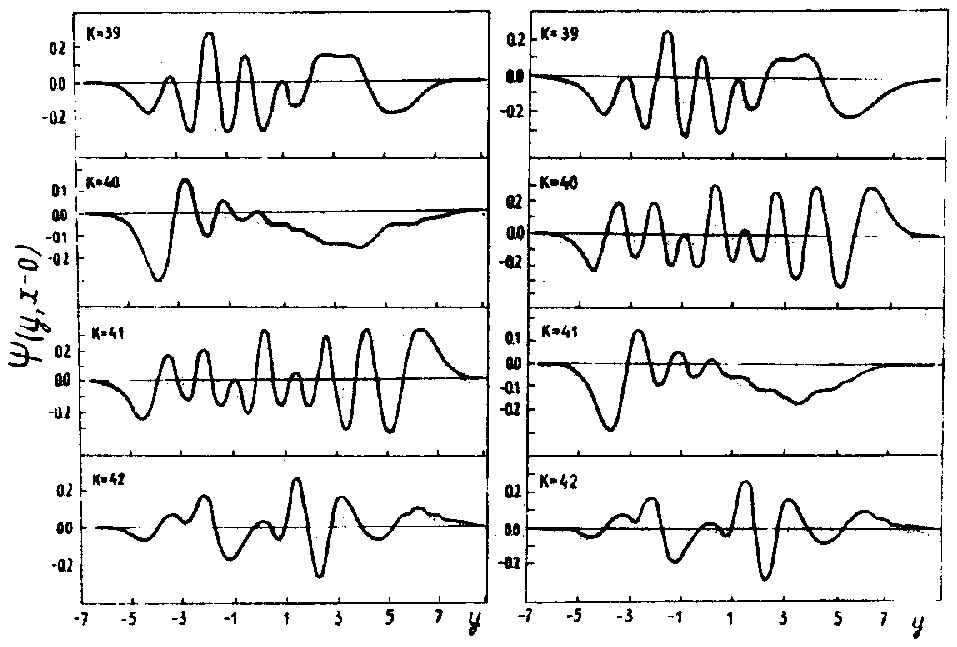}
\caption{Wave functions $\psi \left( {x = 0,y} \right)$ for the
states with $k = 39,40,41,42$ in the neighbourhood of
quasicrossing of the states with $k = 40$ and $41$ . Left $b < b^
*  $ , right $b > b^ *  $ (point of quasicrossing: $b^ *   =
0.098;\;c^ *   = 0.00074$ ). } \label{Fig.30}
\end{figure}

We trace, in particular, the variations of functions with $ K = 40
$ and $K = 41$ , which undergo quasicrossing in the neighbourhood
$ b^ *   \approx 9.8 \cdot \;10^{ - 2} ,\quad c^ *   \approx 7.4
\cdot 10^{ - 4} $ (recall, that under this term we understand the
approaching of levels up to distances of an order of precision of
numerical calculations). The sections of $ \Psi (x = 0,y) $ and
coefficients of expansions $ C_{NLj}^k $ of wave functions of
considering states before and after quasicrossing point are
represented in Figs 30 and 31. The similar characteristics of the
states with $K = 39$ and $K = 42$ which do not undergo
quasicrossing at this values of parameters are represented for
comparison at the same figure.
\begin{figure}[t]
\centering
\includegraphics[height=11cm]{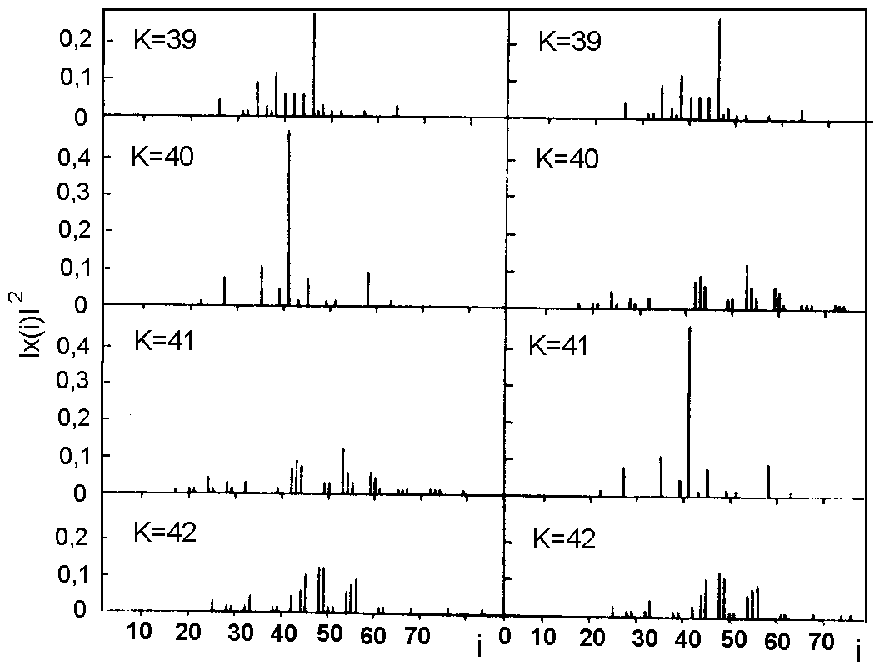}
\caption{Same as Figure 30, but for the coefficients of the
expansion of wave functions $ \left| {x\left( i \right)} \right|^2
= \left| {C_{NLj}^{\left( k \right)} } \right|,\;i $ , - number of
basis state.} \label{Fig.31}
\end{figure}

As it is seen in represented figures the wave functions are not
practically changed for the states 39 and 42, at the same time the
exchange of wave functions is observed for the states $40$ and
$41$ in correspondence with test described above. Analogous
situation takes place for other quasicrossings .

\subsection{ Wave packet dynamics}

The investigation of the time evolution of nonstationary states,
i.e. of packets in quantum systems classical analogs of which
admit chaotic behaviour, gives an important information about
QMCS. The localized quantum wave packet (QWP) is the closest
analog of the point in phase space, which describes the state of
classical system. However, the correspondence between localized
QWP and classical point particle in chaotic region is broken down
in a very short time interval. Let us explain it by using
Takahashi arguments \cite{95}. We take two localized QWP $ \Psi _1
\left( x \right) $ and $ \Psi _2 \left( x \right) $ which are put
in the chaotic region at initial time to be slightly different
from each other so that the difference between $ \left\langle
{\Psi _1 } \right|\hat x\left| {\Psi _1 } \right\rangle $ and $
\left\langle {\Psi _2 } \right|\hat x\left| {\Psi _2 }
\right\rangle $ (or $ \left\langle {\Psi _1 } \right|\hat p\left|
{\Psi _1 } \right\rangle $ and $ \left\langle {\Psi _2 }
\right|\hat p\left| {\Psi _2 } \right\rangle $ ) is very small. We
assume that in chaotic region the localized QWP does not either
extend in a certain time interval of the order $
{\raise0.7ex\hbox{$1$} \!\mathord{\left/ {\vphantom {1 {\sqrt
\hbar  }}}\right.\kern-\nulldelimiterspace}
\!\lower0.7ex\hbox{${\sqrt \hbar  }$}} $ like that in regular
region. Thence, following the Ehrenfest theorem, the packets $\Psi
_1 $ and $\Psi _2 $ move as classical particles and the distance
between $ \left\langle {\Psi _1 } \right|\hat x\left| {\Psi _1 }
\right\rangle $ and $ \left\langle {\Psi _2 } \right|\hat x\left|
{\Psi _2 } \right\rangle $ (or $ \left\langle {\Psi _1 }
\right|\hat p\left| {\Psi _1 } \right\rangle $ and $ \left\langle
{\Psi _2 } \right|\hat p\left| {\Psi _2 } \right\rangle $ ) is
increasing exponentially in time. Let us consider the
superposition $\Psi  = \Psi _1  + \Psi _2 $ which also becomes a
localized QWP in initial state. From the assumption, we can expect
that QWP $\Psi $ does not extend in a certain time interval of the
order $ {\raise0.7ex\hbox{$1$} \!\mathord{\left/ {\vphantom {1
{\sqrt \hbar  }}}\right.\kern-\nulldelimiterspace}
\!\lower0.7ex\hbox{${\sqrt \hbar  }$}} $ . However, considering
the exponential increment of the distance between $ \left\langle
{\Psi _1 } \right|\hat x\left| {\Psi _1 } \right\rangle $ and $
\left\langle {\Psi _2 } \right|\hat x\left| {\Psi _2 }
\right\rangle $ , $\Psi $ extends exponentially in chaotic region
and does not behave as a classical particle. This result is
inconsistent with initial assumption and implies that in chaotic
region the localized QWPs (i.e. $\Psi _1 $ , $\Psi _2 $ and $\Psi
$ ) extend exponentially like a classical probability distribution
in the first stage of time development.

In order to describe such unusual behaviour of QWP one should
address to the concept of a quantum-mechanical phase space. There
are a few well-established schemes to introduce phase-space
variables in quantum mechanics \cite{82,96,97,98}. In the present
study we shall follow the procedure proposed by Weissman and
Jortner \cite{99}. Let us consider an initially localized QWP
$\Psi $ characterized by the coordinates $q$ and the momenta $p$ ,
\begin{equation}\label{3.7.1}
\begin{gathered}
  q = \left\langle \Psi  \right|\hat q\left| \Psi  \right\rangle  \hfill \\
  p = \left\langle \Psi  \right|\hat p\left| \Psi  \right\rangle  \hfill \\
\end{gathered}
\end{equation}
Now we introduce the coherent states $ \left| {p,q} \right\rangle
$ \cite{97}, which in the coordinate x-representa\-tion, are given
by Gaussian wave packets
\begin{equation}\label{3.7.2}
\left\langle {x}
 \mathrel{\left | {\vphantom {x {p,q}}}
 \right. \kern-\nulldelimiterspace}
 {{p,q}} \right\rangle  = \prod\limits_{j = 1}^N {\left( {\pi \sigma _j^2 } \right)^{ - {\raise0.7ex\hbox{$1$} \!\mathord{\left/
 {\vphantom {1 4}}\right.\kern-\nulldelimiterspace}
\!\lower0.7ex\hbox{$4$}}} \exp \left[ { - \frac{{\left( {x_j  -
q_j } \right)^2 }} {{2\sigma _j^2 }} + \frac{{ip_j x_j }} {\hbar }
- \frac{{ip_j q_j }} {{2\hbar }}} \right]}
\end{equation}

In the study of a system of $N$ coupled harmonic oscillators, it
is convenient to choose for constants $\sigma _j $ the rms
zero-point displacements
\begin{equation}\label{3.7.3}
  \sigma _j  = \left( {{\raise0.7ex\hbox{$\hbar $} \!\mathord{\left/
 {\vphantom {\hbar  {m_j \omega _j }}}\right.\kern-\nulldelimiterspace}
\!\lower0.7ex\hbox{${m_j \omega _j }$}}}
\right)^{{\raise0.7ex\hbox{$1$} \!\mathord{\left/
 {\vphantom {1 2}}\right.\kern-\nulldelimiterspace}
\!\lower0.7ex\hbox{$2$}}}
\end{equation}
where mj are masses and $\omega _j $ are frequencies of the
uncoupled oscillators. With this choice of the $\sigma _j $ , the
coherent states $ \left| \alpha  \right\rangle  \equiv \left|
{p,q} \right\rangle $ become the eigenstates of the harmonic
oscillator annihilation operators $a_j $
\begin{equation}\label{3.7.4}
a_j \left| \alpha  \right\rangle  = \alpha _j \left| \alpha
\right\rangle
\end{equation}
where a complex variable $\alpha _j $ ,
\begin{equation}\label{3.7.5}
\alpha _j  = \frac{1} {{\sqrt 2 }}\left( {\frac{{q_j }} {{\sigma
_j }} + i\frac{{\sigma _j }} {\hbar }p_j } \right)
\end{equation}

Using these coherent states, it is possible to introduce the
following quantum - mechanical phase-space density
\begin{equation}\label{3.7.6}
\rho _\Psi  \left( {q,p} \right) = \left| {\left\langle {\alpha }
 \mathrel{\left | {\vphantom {\alpha  \Psi }}
 \right. \kern-\nulldelimiterspace}
 {\Psi } \right\rangle } \right|^2
\end{equation}
where $\Psi $ is any general wave packet. This quantum-mechanical,
coherent-state, phase-space density may be regarded as a quantum
analog of the classical phase-space density, since it satisfies an
equation of motion whose leading term (when expanded in powers of
$ \hbar $ ), corresponds to the classical Liuville equation
\cite{109}. The stationary phase space densities $\rho _E \left(
{p,q} \right)$ are the squares of the projections of the
eigenstates on the coherent state,
\begin{equation}\label{3.7.7}
\rho _E \left( {p,q} \right) = \left| {\left\langle {\alpha }
 \mathrel{\left | {\vphantom {\alpha  E}}
 \right. \kern-\nulldelimiterspace}
 {E} \right\rangle } \right|^2
\end{equation}
where $ \left| \alpha  \right\rangle $ is given in terms of
Gaussian wave packet (\ref{3.7.2}) while the eigenstate $ \left| E
\right\rangle $ is given by Eq.(\ref{3.2.4}). Using the well-known
expressions for scalar products $ \left\langle {\alpha }
\mathrel{\left | {\vphantom {\alpha  {NL}}} \right.
\kern-\nulldelimiterspace} {{NL}} \right\rangle $ \cite{97} we
finally obtain
\begin{equation}\label{3.7.8}
 \begin{gathered}
  \rho _E \left( {p,q} \right) = \frac{1}
{2}\exp \left[ { - \frac{1} {2}\left( {\left| {\alpha _ +  }
\right|^2  + \left| {\alpha _
 -  } \right|^2 } \right)} \right] \cdot  \hfill \\
   \times \sum\limits_{N,L} {\frac{{C_{NL} }}
{{\left( {n_ -  !\;n_ +  !} \right)^{{\raise0.7ex\hbox{$1$}
\!\mathord{\left/
 {\vphantom {1 2}}\right.\kern-\nulldelimiterspace}
\!\lower0.7ex\hbox{$2$}}} }}\left( {\alpha _ + ^{ * n_ +  \;}
\alpha _ - ^{ * n_ -  }  + \;j\;\alpha _ + ^{ * n_ -  } \alpha _
 - ^{ * n_ +  } } \right)\;}  \hfill \\
\end{gathered}
\end{equation}
where $ \alpha _ \pm   = \frac{1} {{\sqrt 2 }}\left( {\alpha _2
\mp i\alpha _1 } \right)$ with $ \alpha _{1,2}  = \frac{1} {{\sqrt
2 }}\left( {q_{1,2}  + ip_{1,2} } \right)$ , $ n_ +   = \frac{{N
+L}} {2} $ , $ n_ -   = \frac{{N - L}} {2}$ . The phase-space
density $\rho _E \left( {p,q} \right)$ is a function of the four
real variables $p_1 $ , $q_1 $ and $p_2 $ , $q_2 $ . We can get
the contour maps of $\rho _E \left( {p_1 ,q_1 ,p_2 ,q_2 } \right)$
in the $q_2  \times p_2 $ plane, taking $q_1  = 0$ and calculating
$p_1 $ from the relation
\begin{equation}\label{3.7.9}
H\left( {q_1  = 0,p_1 ,q_2 ,p_2 } \right) = E
\end{equation}

Quantum Poincare maps (QPM) \cite{99}, obtained by such way,
constitute the quantum analogs of the classical Poincare maps and
can be used for the search of QMCS both in the structure of wave
functions of stationary states and in the dynamics of wave
packets.

Now we shall consider the time evolution of a wave packet which is
initially in a coherent state
\begin{equation}\label{3.7.10}
\left| {\Psi \left( {t = 0} \right)} \right\rangle  = \left|
\alpha  \right\rangle
\end{equation}
The time evolution of such an initially coherent wave packet is
given by
\begin{equation}\label{3.7.11}
 \left| {\Psi \left( t \right)} \right\rangle  = \sum\limits_k {\left|
 {E_k } \right\rangle \left\langle {{E_k }}
 \mathrel{\left | {\vphantom {{E_k } \alpha }}
 \right. \kern-\nulldelimiterspace}
 {\alpha } \right\rangle e^{ - iE_k t} }
\end{equation}
The survival probability$p\left( t \right)$ of finding the system
in its initial state is
\begin{equation}\label{3.7.12}
  p\left( t \right) = \left| {g_\alpha  \left( t \right)} \right|^2
\end{equation}
where $g_\alpha  \left( t \right)$ is the overlap of $\Psi \left(
t \right)$ with the initial state
\begin{equation}\label{3.7.13}
g_\alpha  \left( t \right) = \left\langle {\alpha }
 \mathrel{\left | {\vphantom {\alpha  {\Psi \left( t \right)}}}
 \right. \kern-\nulldelimiterspace}
 {{\Psi \left( t \right)}} \right\rangle  = \sum\limits_k {\left|
 {\left\langle {{E_k }}
 \mathrel{\left | {\vphantom {{E_k } \alpha }}
 \right. \kern-\nulldelimiterspace}
 {\alpha } \right\rangle } \right|^2 e^{ - iE_k t} }
\end{equation}
Utilizing the definition of the stationary phase-space density
(\ref{3.7.7}), we can write
\begin{equation}\label{3.7.14}
g_\alpha  \left( t \right) = \sum\limits_k {\rho _{E_k } e^{ -
iE_k t} }
\end{equation}
Equation (\ref{3.7.14}) implies that dynamics is determinated by
the spectrum of the initial coherent state $\left| \alpha
\right\rangle $ .

Weissman and Jortner \cite{99} have observed for Henon-Heiles
Hamiltonian (QON Hamiltonian with $m = 1;a = 1;c = 0$ ) two
limiting types of QWP dynamics of initially coherent Gaussian wave
packets, which correspond to quasiperiodic time evolution and to
rapid decay of the initial state population probability.
Quasiperiodic time evolution is exhibited by wave packets
initially located in regular region, while rapid decay of the
initial state population probability is revealed by those wave
packets that are initially placed in irregular regions.

Ben-Tal and Moiseyev \cite{108} calculated the survival
probability $p\left( t \right)$ for an initial complex Gaussian
wave packets by the Lanczos recursion method. This method has a
practical value since it requires $nN^2 $ numerical operations
($n$ is number of Lanczos recursion, $N$ is dimension of the
original Hamiltonian matrix) rather than $N^3 $ required to
calculate the eigenvectors $H$ . For bounded time calculations n
is much smaller than $N$ .

Let us turn to the consideration of the dynamics of QWP in the PES
with some local minima $\left( {W > 16} \right)$ . Transitions
between different local minima can be divided into induced (the
excitation energy exceeds the value of potential barrier) and
tunnel transitions. The latter ones are subdivided into
transitions from discrete spectrum into continuous spectrum (for
example, $\alpha $ -decay, spontaneous division) and from discrete
spectrum into discrete spectrum (for example, transitions between
isomeric states). The process of tunneling across a
multidimensional potential barrier, when initial and final states
are in discrete spectrum, is the most complicated and almost
noninvestigated problem.

The time evolution of wave packet is more often studied by two
methods \cite{100}: either by direct numerical integration of the
Schrodinger time-dependent equation with corresponding initial
condition $\Psi \left( {\vec r,t = 0} \right)$ , or by expansion
of the packet $\Psi \left( {\vec r,t} \right)$ in the
eigenfunctions of the stationary problem. The first method has
some lacks, e.g., the complication of interpretation of the
obtained results and the necessity of separating the contributions
from subbarrier and tunnel transitions for the packet of an
arbitrary shape. These difficulties can be avoided providing that
the subbarrier part of the spectrum $E_n (E_n  < U_0 ,U_0 $ -the
height of barrier) and the corresponding stationary wave functions
$\psi _n \left( r \right)$ are known. A pure tunnel dynamics will
take place for the packets representable in the form
\begin{equation}\label{3.7.15}
\Psi \left( {\vec r,t} \right) = \sum\limits_n {C_n e^{ - \frac{i}
{\hbar }E_n t} \psi _n \left( {\vec r} \right),E_n  < U_0 }
\end{equation}
where
\begin{equation}\label{3.7.16}
 C_n  = \int {\Psi \left( {\vec r,t = 0} \right)\psi _n \left(
  {\vec r} \right)d\vec r}
\end{equation}
The probability $p^R \left( t \right)$ of finding the particle at
the moment of time t in certain local minimum $R$ is
\begin{equation}\label{3.7.17}
p^R \left( t \right) = \int\limits_R {\left| {\Psi \left( {\vec
r,t} \right)} \right|^2 d\vec r = \sum\limits_{m,n} {C_m^ *  C_n
e^{\frac{{i\left( {E_m  - E_n } \right)}} {\hbar }} \int {\psi _m^
*  \left( {\vec r} \right)\psi _n \left( {\vec r} \right)d\vec r}
} }
\end{equation}
or in the two level approximation
\begin{equation}\label{3.7.18}
 p^R \left( t \right) = p^R \left( 0 \right) - 4C_1 C_2
 \sin ^2 \left( {\frac{{\frac{1}
{2}\left( {E_1  - E_2 } \right)t}} {\hbar }} \right)\int {\psi _1^
*  \left( {\vec r} \right)\psi _2 } \left( {\vec r} \right)d\vec r
\end{equation}
Let us introduce \cite{101} the value
\begin{equation}\label{3.7.19}
\bar p^R  = \max _{\forall t} p^R \left( t \right)
\end{equation}
which is the maximum probability of finding the particle  in the
certain local minimum $R$ , if initially it was localized in the
arbitrary minimum. If the number of local minima is more than two,
then of independent interest is
\begin{equation}\label{3.7.20}
  \bar{ \bar{ p}}^{R_0 }  = \min _{\forall t} p^{R_0 } \left( t \right)
\end{equation}
i.e., minimum probability of finding the wave packet in the
minimum $R_0 $ corresponding to its initial localization.
\begin{figure}
\centering
\includegraphics[height=6cm]{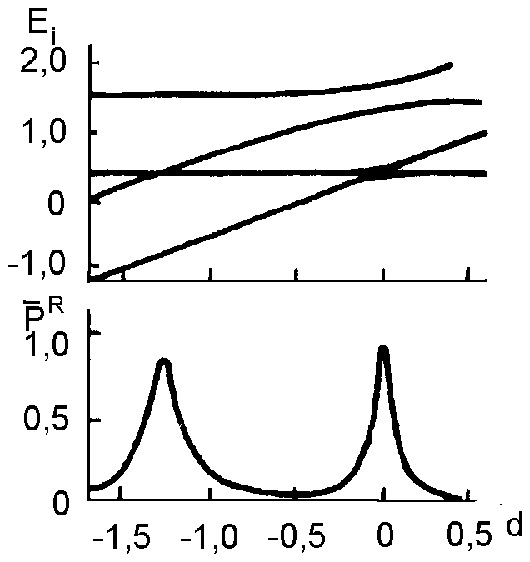}
\caption{Dependence of subbarier energy levels $E_i $ for double
asymmetric one-dimensional rectangular well with infinite external
walls (a) and $ \bar p^R $ , (b) from $d$ . Width of well is equal
to $3$ , width of barrier - $1$ , height of barrier - $2$ .}
\label{Fig.32}
\end{figure}

Intuitively, we may suggest that $ \bar p^R  \approx 1 $ if the
initial minimum is local, and the final one is absolute. However,
the results \cite{101} obtained for the simplest one-dimensional
models (asymmetric double wells of different shapes) are
inconsistent with the intuitive expectations. The probability of
tunneling from the local minimum to the absolute one resonantly
depends on the potential parameters. Fig.32  gives the dependence
of $ \bar p^R $ from well depth displacement d. It is seen that at
an arbitrary asymmetry $\bar p^R  \ll 1$ .
\begin{figure}
\centering
\includegraphics[height=6cm]{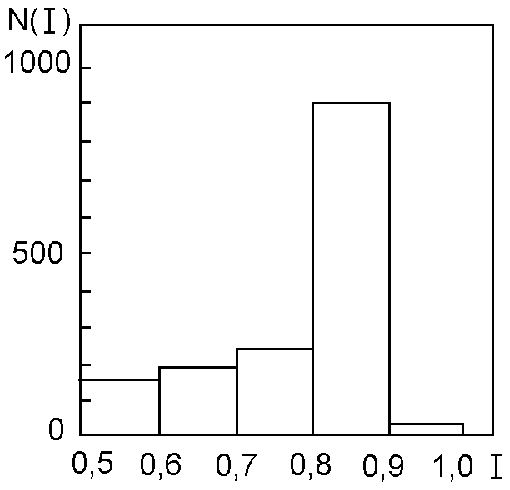}
\caption{Distribution of subbarier wave functions $N\left( I
\right)$ .} \label{Fig.33}
\end{figure}

The resonant behaviour of $\bar p^R $ becomes clear if one
considers the spatial structure of the subbarrier wave functions.
For a sufficiently wide barrier at an arbitrary asymmetry, the
subbarrier wave functions are largely localized in separate
minima. Fig.33 gives the distribution $N\left( I \right)$ of the
subbarrier wave functions as their degree of localization $ I_n =
\int\limits_R {\left| {\psi _n \left( {\vec r} \right)} \right|^2
d\vec r} $ for an accidental choice of the well depth displacement
$d$. The delocalization takes place only in the vicinity of the
quasicrossing of level. The degree of this delocalization is
directly depends on the distance between the interacting levels.
Obviously, the QWP, in which the components localized in the
certain minimum are dominating, cannot effectively tunnel to the
neighbouring minimum. This just explains the stringent correlation
between the $\bar p^R $ minima  and the level quasicrossing (see
Fig. 32.)

Now the question arises if a similar correlation between the level
quasicrossings, the delocalization of wave functions and resonant
tunneling persists in the two-dimensional case. To give an answer
to this question let us turn to the results of numerical
investigation of stationary wave functions of Hamiltonian QON in
region $W > 16$ \cite{102}. Recall that three ($C_{3v} $ -
symmetry) identical additional minima appear at $W > 16$ apart
from the central minimum. The central minimum exceeds the lateral
ones in depth in the region $16 < W < 18$. At $W > 18$ , the
central minimum becomes the local one. In this region of
parameters the procedure of diagonalization of Hamiltonian QON in
oscillator basis essentially  complicates. It is connected with
the fact, that the basis of Hamiltonian, the potential of which
has the unique minimum, is used for the diagonalization of
Hamiltonian possessing complex topology of the PES.
\begin{figure}
\centering
\includegraphics[height=6cm]{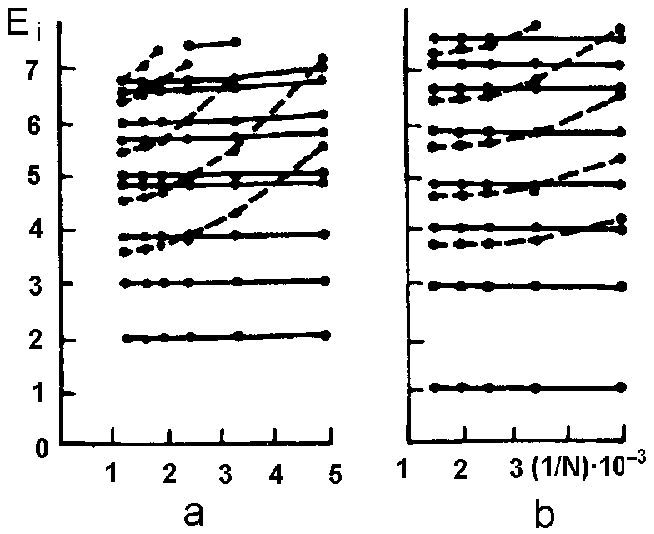}
\caption{Dependence of energy spectrum $E$ -type (a) at $b =
0.173798$ and $A_1 $ -type (b) at $b = 0.17$ from the dimension of
submatrices $N$ for $W = 17.8$ . Dashed lines correspond to the
states localized in peripheral minima. } \label{Fig.34}
\end{figure}
\begin{figure}
\centering
\includegraphics[height=6cm]{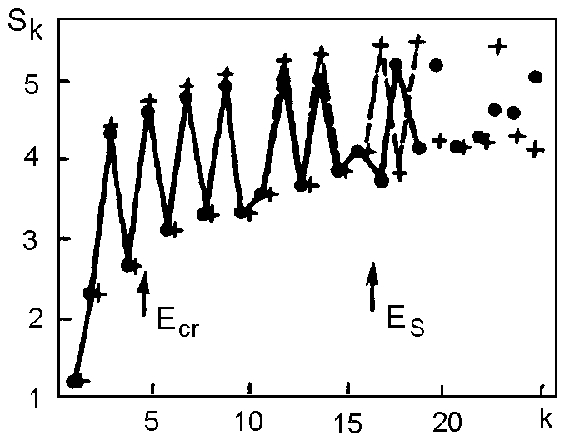}
\caption{Dependence of $S_k $ from number of eigenstate $k$ ($A_1
$ - type) for the dimension of submatrices 408 (dark point) and
690 (crosses) at $W = 17.8$ and $b = 0.17$ ; $E_S $ corresponds to
saddle energy and $E_{cr} $ - the classical critical energy of the
transition from regular type of motion to chaotic one.}
\label{Fig.35}
\end{figure}

In addition to large dimension of the basis it is necessary that
the basis wave functions have sufficient value in the region of
lateral minima. It is achieved by the optimization of oscillator
frequency of basis $\omega _0 $ . For the values of parameters
used in the calculations $\left( {W = 17.8;b = 0.17} \right)$ the
value of oscillator frequency $\omega _0  = 0.2$ . Fig.34 gives
low eigenvalues of $E$ and $A_1 $ types depending on the basis
dimension. We can see, that it is possible to get the saturation
in basis at the dimension of submatrices $N \sim 10^3 $ even for
low states localized in the lateral minima (dashed lines).
\begin{figure}
\centering
\includegraphics[height=8cm]{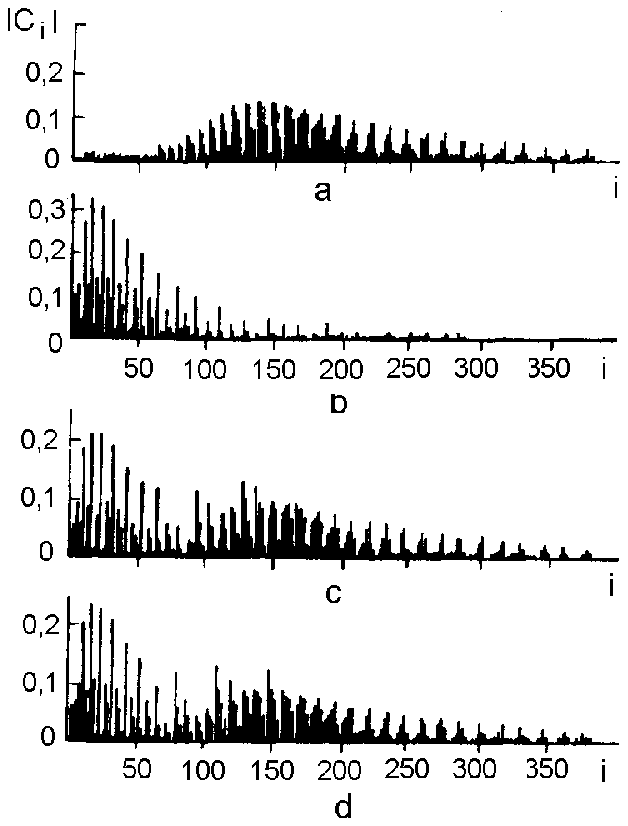}
\caption{Distribution of coefficients $\left| {C_i } \right|$ by
number of basis state $i = \{ N,L,j\} $ ($E$ - type) for the
localized states with $k = 3$ (a) and $k = 4$ (b) at $b =
0.17379\;\left( {W = 17.8} \right)$ and for the delocalized states
with $k = 3$ (c) and $k = 4$ (d) in the point of quasicrossing $b
= 0.1737924\;\left( {W = 17.8} \right)$ .} \label{Fig.36}
\end{figure}

We can also use introduced above the analog of thermodynamic
entropy $S_k $ for estimation of degree of saturation in basis.
Fig.35 gives the $S_k $ values for the states of $A_1 $ type for
the dimension of submatrices 408 and 690. Increasing of the basis
does not lead to sufficient changes of values $S_k $ for the
states with energy up to saddle energy $E_S $ .

The states,localized in the central or in the lateral minima, have
essentially different distributivity of coefficients $C_i^k \quad
\left\{ {i = NLj} \right\}$(see Fig. 36a,b) and thus different
entropies:  states, localized in the central minimum, have less
entropy). In the neighbourhood of the points of level
quasicrossings, the delocalization of wave functions,
corresponding to these levels, takes place; these wave functions
possess close distributivity of coefficients $C_i^k $ (see Fig.
36.c.d).
\begin{figure}
\centering
\includegraphics[height=6cm]{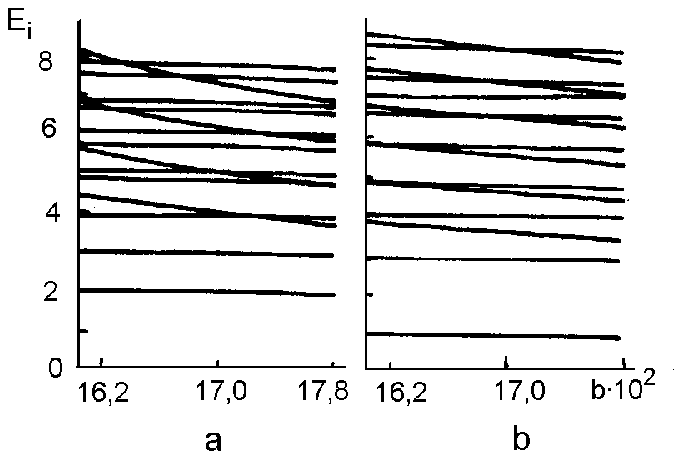}
\caption{Dependence of energy levels $E_i $ of Hamiltonian (16.1)
from the parameter $b$ for $W = 17.8$ : a - spectrum of $E$ -
type, b - spectrum of $A_1 $ - type.} \label{Fig.37}
\end{figure}
Fig.37 shows the subbarrier part of the energy spectrum obtained
by the diagonalization. As it is easy to see the tunneling of the
wave packet, composed of the subbarrier wave functions, can be
described in the two-level approximation. Indeed, there are
approximately 10 level quasicrossings of $A_1 $ and $E$ - types,
where the nonlinearity parameter changes are of the order of $10^{
- 2} $ . The effective half-width of the overlap integral in
(\ref{3.7.18}) is about $10^{ - 5} $ (see Fig. 38).Hence, all
nondiagonal elements will be close to zero with the overwhelming
probability in the matrix $ \alpha _{mn}  = \int\limits_R {\psi m^
*  \left( {\vec r} \right)\psi _n } \left( {\vec r} \right)d\vec r $
for an arbitrary nonlinearity parameter (e.g., $b$ ). Two
appreciable different from zero nondiagonal matrix elements
(two-level approximation) appear only in the vicinity of
quasicrossings. The probability of double quasicrossings at a
fixed nonlinearity parameter is almost excluded. This probability
is by two or three orders lower than that of rather rare $\left(
{~10^{ - 3} } \right)$ single quasicrossings.
\begin{figure}[t]
\centering
\includegraphics[height=8cm]{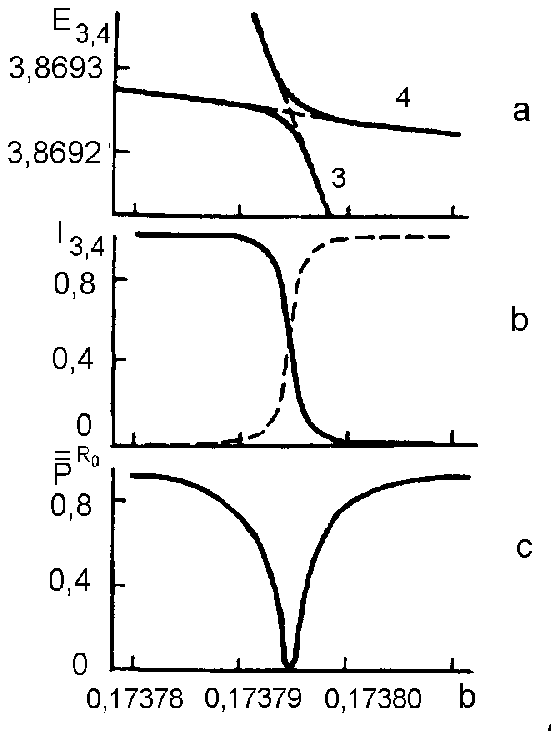}
\caption{ a - quasicrossing of energy levels of $E$ - type with $k
= 3,4$ for Hamiltonian (16.1); b - localization of wave functions
of states with $k = 3,4$ in the central well at different $b$ ; c
- dependence $ \bar p^R $ from $b$ .} \label{Fig.38}
\end{figure}

So, now we can give the answer to the question we have posed
above. The stringiest correlations between level quasicrossings,
delocalization of wave functions and resonant tunneling across the
potential barrier take place in the two-dimensional case (and,
most likely, in the multi-dimensional one too).

The existence of the mixed state for the many-well potentials must
be well manifested itself in dynamics of the QWP. Preexponential
factor of the tunnel amplitude depends on the type of classical
motion and consequently we expect to observe the asymmetry of the
effective barrier penetration in the mixed. This purely classical
effect may be observed only if the uncertainty in level energy is
comparable with the average distance between levels and the system
does not yet "feels" that the spectrum is discrete. This
determines the time scale for which observation of effect is
possible. It is the same time scale on which transition from
classical linear diffusive increasing of energy to quasiperiodical
quantum evolution is observed \cite{103}.

\section{ Summary and open questions}

\subsection{ Summary}

In this review we have presented a complete description of
classical dynamics generated by the Hamiltonian of quadrupole
oscillations and identified those peculiarities of quantum
dynamics which can be interpret as QMCS.

We have shown an intimate connection between dynamics features and
geometry of the PES. Interpretation of negative curvature of the
PES as the source of the local instability allowed to correctly
predict the critical energy of the transition to chaos for
one-well potentials.

Particular attention has been given to investigation of classical
dynamics in the parameter region according to the PES with a few
local minima. As is shown, one of the main peculiarities of the
many-well Hamiltonians is the mixed state: at one at the same
energy in a different local minima diverse dynamical regimes
(regular or chaotic) are realized. Until the present time the most
theoretical and numerical works have been focused on the behaviour
of so-called billiards or Hamiltonian systems with the simplest
topology of the PES. Evidently it is not enough for understanding
of the real many-body systems with the complicated PES for which
the mixed state is a situation of general position. Therefore the
present review should be read as one of the indispensable steps on
the way to transition from description of the model systems to
direct consideration of much more realistic systems.

A new methodology for determination of the critical energy of the
transition to chaos based on the investigation of correlations
between dynamic features and geometry of the PES is developed for
many-well potentials. It is more simpler than criteria of
transition to chaos connected with one or other version of overlap
resonance's criterion. At the energies below the critical one
approximate integrals of motion for every of local minima are
derived by the Birkhoff-Gustavson normal form technique.

We have numerically demonstrated that for potential with a
localized unstable region (in particular localized region of
negative Gaussian curvature) regular motion restores at high
energy, i.e. transition regularity-chaos-regularity takes place
for these potentials.

QMCS in dynamics of QON have been the center of our consideration.
The variations of statistical properties of energy spectrum in the
process of $R - C - R$ transition have been studied in detail. For
the chaotic region all the analyzed statistical characteristics
are seen to be in good agreement with the GOE predictions. For
regular regions our calculations are consistent with the
hypothesis of the universal character of energy spectrum
fluctuations. Though for small level spacings we observed some
deviations which were probably due to a small admixture of the
chaotic component.

We proved that the type of classical motion is correlated with the
structure of the stationary wave functions of highly excited
states in the $R - C - R$ transition. Correlations were found both
in the coordinate space (the lattice of nodal curves and the
distribution of the probability density) and in the Hilbert space
associated with the integrable part of Hamiltonian (the
distribution of the wave functions in the oscillator basis and the
entropy of individual eigenstates). Calculations with the scaled
Planck constant, which make it possible to obtain wave functions
with equal quantum numbers and energies that correspond to
different types of classical motion, enabled us to unambiguously
isolate correlation effects in the structure of wave functions.

The Hamiltonian of QON was used as an example to study the shell
structure destruction induced by the increase of nonintegrable
perturbation term which models residual nucleon-nucleon
interaction. In the vicinity of the classical critical energy
value it was observed multiple formation of quasicrossings of the
energy levels, violation of the quasiperiodical energy dependence
of the entropy, and the increase of the average value fluctuations
of the operators used to classify the eigenstate of the integrable
problem.

The optimization of the basis frequency and, as a consequence, the
possibility of diagonalization matrices of higher dimension made
it possible to trace the restoration of the shell structure in the
transition $C - R_2 $ from chaos to regularity.

\subsection{Open questions}

The problem of studying of QMCS in Hamiltonian systems with a few
local minima represents almost a noninvestigated region. We must
take into account that, apart from conceptual difficulties,
construction of spectrum and eigenfunctions in this case is a very
difficult computational problem. Calculation of energy splitting
$DE$ due to tunneling is one of the oldest problem of quantum
mechanics. One dimensional problem for the most part is
understood. A complication arises in the multidimensional problem,
which might not have been expected until recently. This is because
the semiclassical wave functions which determine energy splitting
are sensitive to the nature of the classical motion. Wilkinson
[104] calculated the energy splitting due to tunneling between a
pair of quantum states which correspond to classical motion on
tori in phase space. The case where both wave functions correspond
to classical chaotic motion were discussed by Wilkinson and Hannay
[105]. Problem of the wave function structure for the mixed state
remains to be solved. The configuration space in this situation
breaks up naturally into different regions (separate local
minima), in each of which different dynamics regimes are realized.
Thus, there is a need of a calculating scheme, which will enable
us to use the partial information about each isolated region to
obtain solution of the full problem. We believe that the path
decomposition expansion of Auerbach and Kivelson \cite{106} will
prove to be useful for solving $2 - d$ tunneling problem in the
mixed state. This formalism allows to express the full time
evolution operator as a time convolution and surface integrations
of products of restricted Green functions, each of which involves
the sum over paths that are limited to different regions of
configuration space. Even for complicated nonseparable potential
the qualitative behaviour is readily inferred and quantitative
solutions can be obtained from knowledge of the classical
dynamics.

In conclusion, we note that an important problem concerning the
role of periodic orbit in the structure of wave functions was not
considered in this study. In the semiclassical limit quantum state
must be determined by the invariant sets formed by classical
trajectories. These sets cover the total energy surface for
systems that are chaotic in the classical limit. The approach in
which the energy surface as a whole plays a dominant role is
confirmed by the chaotic structure of nodal curves and by the
random distribution of the probability density of the wave
functions of highly exited states. However, the recent findings of
Heller \cite{107}, who showed that wave functions can have the
so-called scars corresponding to a high concentration of
probability density near unstable periodic orbits, demonstrate
that such an approach is not complete. The reason for this is
that, although the measure of periodic trajectories is zero, their
contribution is essentially singular, in contrast to the smooth
contribution from the energy surface. These problems require
special analysis for the Hamiltonian under study.

\vspace{.5cm}

\noindent {\bf Acknowledgements} The authors are grateful  to A.
Pashnev, G. Sotnikov and V. Yanovsky for fruitful discussions and
valuable advices. The work was done with a partial support of the
Basic Research Foundation of Ukraine (grant 383).


\section{References}

\vspace{-1cm}

\begin{thebibliography}{99}
\bibitem{1} B.Chirikov,Phys.Rep.{\bf 52} (1979) 263
\bibitem{3} A.J.Lichtenberg and M.A.Liberman,Regular and Stochastic Motion
      (Springer, New York,1983)
\bibitem{4} F.Haake, Quantum Signatures of Chaos (Springer, Berlin,1991)
\bibitem{6}M.C. Gutzwiller, Chaos in Classical and Quantum Mechanics (Springer Verlag, Berlin,
      1991)
\bibitem{7}R.D.Williams and S.E.Koonin, Nucl.Phys.{\bf A391}(1982) 72
\bibitem{12}Yu.L.Bolotin, V.Yu.Gonchar,E.V.Inopin,V.V.Levenko, V.N.Tarasov and\\
       N.A.Chekanov, Fiz.Elem.Chastits and At.Yadra {\bf 20}(1989) 878
\bibitem{13}W.Swiateski, Nucl.Phys.{\bf A488}(1989) 375
\bibitem{22}S.Bjornholm, Nucl.Phys.{\bf A447}(1985) 117
\bibitem{23}T.H.Seligman and H.Nishioka Eds, Quantum Chaos and Statistical Nuclear Physics
       (Springer, Heidelberg, 1986)
\bibitem{53} M.Tabor, Chaos and Integrability in Nonlinear Dynamics (John Wiley, New York,
       1989)
\bibitem{54} M.Berry, in Chaos and Quantum Physics, Les Houches, LII, ed M.Giannoni, A.Voros
       and J.Zinn-Justin (North-Holland, Amsterdam,1991) p.253
\bibitem{55} B. Eckardt, Phys. Rep.{\bf 163} (1988) 205
\bibitem{56} O.Bohigas, S.Tomsovic and D.Ulimo, Phys.Rep.{\bf223}(1993) 43
\bibitem{57}P.V.Elyutin, Usp.Fiz.Nauk{\bf 155}(1988) 397
\bibitem{58} Y.Weissman and J.Jortner, J.Chem.Phys.{\bf 77} (1982) 1469.
\bibitem{59} M.D.Feit, J.A.Fleck,Jr., and A.Steiger, J.Comput.Phys. {\bf 47} (1982) 412.
\bibitem{60} M.D.Feit and J.A.Fleck,Jr., J.Opt.Soc.Amer. {\bf 17} (1981) 1361.
\bibitem{61} J.A.Fleck,Jr., J.R.Morris and M.D.Feit, Appl.Phys. {\bf 10} (1976) 129.
\bibitem{62} N.A.Chekanov, Yad.Fiz. {\bf 50} (1989) 344.
\bibitem{63} M.Robnik, J.Phys.A: Math.Gen. {\bf 17} (1984) 109.
\bibitem{64} T.Uzer and R.A.Marcus, J.Chem.Phys. {\bf 81} (1984) 5013.
\bibitem{65} T.A.Brody, J.Flores, J.B.French, P.A.Mello, A.Pandey and S.S.Wong\\
       Rev.Mod.Phys. {\bf 53} (1981) 385.
\bibitem{66} L.D.Landau and E.M.Lifshitz, Quantum Mechanics (Pergamon, New York, 1977).
\bibitem{67} C.E.Porter (ed) Statistical Theory of Spectra: Fluctuations
       (Academic, New York, 1965).
\bibitem{68} O.Bohigas, M.Giannoni and C.Shmit, Phys.Rev.Lett. {\bf 52} (1983) 1.
\bibitem{69} T.Seligman, J.Verbaarshot and M.Zirnbauer, Phys.Rev.Lett. {\bf 53} (1984) 215.
\bibitem{70} O.Bohigas, R.V.Hag and A.Pandey, Phys.Rev.Lett. {\bf 54} (1985) 645.
\bibitem{71} D.Delande and J.Gay, Phys.Rev.Lett. {\bf 57} (1986) 2006.
\bibitem{72} M.Berry and M.Tabor, Proc.R.Soc. {\bf A356} (1977) 375.
\bibitem{73} M.Berry and M.Robnik, J.Phys.A: Math.Gen.{\bf 17} (1984) 2413.
\bibitem{74} F.M.Izrailev, Phys.Rep. {\bf 196} (1990) 299.
\bibitem{75} F.J.Dyson and M.L.Mehta, J.Math.Phys. {\bf 4} (1963) 701.
\bibitem{76} E.B.Bogomolny,JETP Lett. {\bf 41} (1985) 55.
\bibitem{77} S.S.M.Wong , in: Chaotic Behavior in Quantum Systems, ed. G.Casati
      (Plenum Press, New York, 1985).
\bibitem{78} Yu.L.Bolotin, V.Yu.Gonchar, V.N.Tarasov and N.A.Chekanov,
       Phys.Lett. {\bf A135} (1989) 29.
\bibitem{79} Y.Ersin, Phys.Rev. {\bf A38} (1988) 1027.
\bibitem{80} D.C.Meredith, S.E.Koonin and M.R.Zirnbauer, Phys.Rev. {\bf A37} (1988) 3499.
\bibitem{81} H.J.Lipkin, N.Meshkov and A.J.Glick, Nucl.Phys. {\bf 62} (1965) 188.
\bibitem{82} E.P.Wigner, Phys.Rev. {\bf 40} (1932) 749.
\bibitem{83} M.V.Berry, J.Phys.A: Math.Gen. {\bf 10} (1977) 2083.
\bibitem{84} S.W.McDonald and A.N.Kaufman, Phys.Rev. {\bf A37} (1988) 3067.
\bibitem{85} Yu.L.Bolotin, V.Yu.Gonchar, V.N.Tarasov and N.A.Chekanov,
       Phys.Lett. {\bf A144} (1990) 459.
\bibitem{86} Yu.L.Bolotin,V.Yu.Gonchar and V.N.Tarasov, Yad.Fiz. {\bf 58} (1995) 1499.
\bibitem{87} R.M.Stratt,C.N.Handy and W.H.Miller, J.Chem.Phys. {\bf 71} (1979) 3311.
\bibitem{88} K.S.J.Nordholm and S.A.Rice, J.Chem.Phys. {\bf 61} (1974) 203.
\bibitem{89} F.Yonezava, J.Non-Cryst.Solids {\bf 35} (1980) 26.
\bibitem{90} L.E.Reichl, Europhys.Lett. {\bf 6} (1988) 669.
\bibitem{91} J.Keller and S.Rubinov, Ann.Phys. {\bf 9}(1960) 24.
\bibitem{92} I.S.Persival, Adv.Chem.Phys. {\bf 36} (1977) 1.
\bibitem{93} M.V.Berry and M.Wilkinson, Proc.R.Soc. {\bf A392} (1984) 15.
\bibitem{94} J.M.Eisenberg and W.Greiner, Nuclear Theory,
      Vol.1 (North-Holland, Amsterdam,(1970)
\bibitem{95} K.Takahashi, Prog.Theor.Phys. Supplement {\bf 98} (1989) 109.
\bibitem{96} J.Zak, Phys.Rev. {\bf 177} (1969) 1151.
\bibitem{97} R.J.Glauber, Phys.Rev. {\bf 131} (1963) 2766.
\bibitem{98} J.R.Klauder and E.C.Sudarshan, Fundamentals of Quantum Optics
       (Benjamin, New York, 1968).
\bibitem{99} Y.Weissman and J.Jortner, J.Chem.Phys. {\bf 77} (1982) 1486.
\bibitem{100} M.Razavy and A.Pimpale, Phys.Rep. {\bf 168} (1988) 307.
\bibitem{101}M.M.Nieto, V.P.Gutschik, C.M.Bender, F.Cooper and D.Strottman,
       Phys.Lett. {\bf B163} (1985) 336.
\bibitem{102} Yu.L.Bolotin, V.Yu.Gonchar and V.N.Tarasov, Ukr.Fiz.J. {\bf 38}
(1993) 513.
\bibitem{103} G.Casati, B.V.Chirikov, F.M.Izrailev and J.Ford, in:
Stochastic Behavior in Classical and Quatum Hamiltonian Systems,
eds J.Casati and J.Ford (Springer-Verlag,Berlin,1979).
\bibitem{104} M.Wilkinson, Physica {\bf D21} (1986) 341.
\bibitem{105} M.Wilkinson and J.H.Hannay, Physica {\bf D27} (1987) 201.
\bibitem{106} A.Auerbach and S.Kivelson, Nucl.Phys. {\bf B257} (1985) 799.
\bibitem{107} E.Heller, Phys.Rev.Lett. {\bf 53} (1984) 1515.
\bibitem{108} N.Ben-Tal and and N.Moiseev, J.Phys.A:Math.Gen. {\bf 24} (1991)
3593.
\bibitem{109} E.Prugovecki, Ann.Phys.{\bf 110} (1978) 102.
\bibitem{110} O.Bohigas, M.Giannoni, Lect. Notes in Phys. {\bf 209} (1984) 1.

\end{thebibliography}
\end{document}